\documentstyle[aps,eqsecnum,epsfig]{revtex}

\widetext

\begin{document}
\title{Non-Markovian quantum feedback from homodyne measurements: the 
effect of a non-zero feedback delay time}
\author{V. Giovannetti, P. Tombesi and D. Vitali}
\address {Dipartimento di Matematica e Fisica, Universit\`a di
Camerino, via Madonna delle Carceri I-62032 Camerino \\
and Istituto Nazionale per la Fisica della Materia, Camerino, Italy}
\date{\today}
\maketitle

\begin{abstract}
We solve exactly the non-Markovian
dynamics of a cavity mode in the presence of a feedback 
loop based on homodyne measurements, in the case
of a non-zero feedback delay time. With an appropriate choice
of the feedback parameters, this scheme is able to
significantly increase the decoherence time of the cavity mode,
even for delay times not much smaller than the decoherence time itself.
\end{abstract}


\section{Introduction}

Although feedback schemes have been used for long times to control noise,
a general theory of feedback for quantum systems has been developed
only some years ago by Wiseman and Milburn \cite{wis1,wis2,wis3}.
Interesting possibilities are
opened by the ability to control systems at the quantum
level using appropriate feedback loops
and some of them have been showed in a series
of papers \cite{squee,qnd,GTV,prl,opto,pran}. Ref.~\cite{squee}
has shown that an electro-optical feedback loop based on homodyne 
measurements of a cavity mode provides an affordable way
to realize a squeezed bath for the mode. As a
consequence,
homodyne-mediated feedback can be used to get squeezing \cite{qnd}, and,
in the case of optical cavities with an oscillating mirror, 
it can be used to significantly cool the mirror. This fact can be
extremely useful for the interferometric detection of gravitational waves
\cite{opto}. The application of a feedback loop realizes an effective 
``reservoir engineering'' \cite{poy} and therefore it can be useful 
also for decoherence control, which is a rapidly expanding field since
decoherence is the main limiting factor for quantum information processing
\cite{eke}. Refs.~\cite{qnd,GTV,prl,pran} have already shown that 
the decoherence induced by photon leakage in electromagnetic
cavities can be significantly 
suppressed with appropriate feedback loops, using the homodyne
photocurrent in \cite{qnd,GTV} and direct photodetection and atomic injection
in \cite{prl,pran}.

However, all the relevant applications considered up to now always assume 
the zero feedback delay time limit $\tau \rightarrow 0$, which is much easier
to handle because the problem becomes Markovian and the effect of feedback can
be expressed in terms of an effective master equation
\cite{wis1,wis2,wis3}. The presence of a non-zero delay has been considered
briefly only in \cite{wis3}, where the spectrum of a homodyne measurement
has been evaluated for a simple case. 
The Markovian treatment is justified whenever the
feedback delay time is much smaller than the typical cavity timescale. If one 
considers squeezing or some other stationary state phenomenon, the feedback 
delay time $\tau$ has to be compared with the cavity relaxation 
time $\gamma^{-1}$
and for sufficiently good cavities the Markovian condition
$\gamma \tau \ll 1$ is usually satisfied. 
However, if one considers the feedback 
scheme for decoherence control, the delay $\tau$ has to be negligible with 
respect to the decoherence time $t_{dec} \simeq (\gamma \bar{n})^{-1}$,
which can be much shorter than the damping time when the cavity
mean photon number $\bar{n}$ is large. In these cases,
the unavoidable non-zero feedback delay time may have important effects
and it would be important to deal with the exact non-Markovian problem
with $\tau \neq 0$. There is in fact a renewed interest in 
non-Markovian effects, which can play an important role when 
considering quantum optics in high-Q cavities and in photonic bandgap 
materials. For this reason, non-Markovian trajectory theories have 
been recently developed in Refs.~\cite{molm,diosi,jack}.

The quantum theory of feedback has been developed by Wiseman and Milburn
in \cite{wis1,wis2} using quantum trajectory theory \cite{carm}, 
and only later Wiseman showed an equivalent derivation based on the 
input-output theory \cite{gar0,gar1,gar2}
in Ref.~\cite{wis3}. However Ref.~\cite{wis3} proved the equivalence
between the two approaches in the perfect 
detection $\eta =1$ case only.
In this paper we shall see how to extend the quantum Langevin approach
of the input-output theory to the non-unit efficiency case
and we shall see that this theoretical framework is best suited to deal with 
the non-Markovian case of non-zero feedback delay time. 
We shall consider the non-Markovian effects by completely solving
the dynamics of a cavity mode in the presence of a homodyne-mediated
electro-optical feedback loop, which has been already considered 
(in the zero delay limit only) in Ref.~\cite{GTV}.

The paper is organized as follows. In Section II we shall reconsider
the quantum theory of feedback in the case of homodyne measurements,
adopting the input-output theory of Gardiner and Collett \cite{gar0,gar1,gar2}
and we shall see how to introduce the non-unit detection efficiency 
in this framework. In Section III we shall completely solve the non-Markovian
dynamics in the presence of a non-zero feedback delay by considering the time
evolution of the probability distribution of the measured field quadrature
and of the characteristic function.
We shall consider in particular the possibility of inhibiting the
decoherence of a Schr\"odinger cat state initially generated in the cavity
and we shall see that the significant decoherence suppression which
can be obtained, for appropriately chosen feedback parameters, in the zero delay
case (see Ref.~\cite{GTV}) is recovered even in the presence of not negligible
feedback delay times.

\section{Homodyne-mediated quantum feedback theory within the input-output 
formalism}

We shall consider an optical cavity, with annihilation
operator $a$, subject to the homodyne
measurement of the field quadrature
\begin{equation}
    X_{\varphi} = \frac{1}{2} ( \: a \, e^{ - i \varphi} \,+ \, a^{ \dagger} 
    \, e^{  i \varphi} \: ).
\label{qm}
\end{equation}
We shall consider the possibility of applying a feedback loop
to this cavity mode, by feeding back part of the output 
homodyne photocurrent to control in some way the $a$ mode dynamics.

First of all it is convenient to
reformulate Wiseman and Milburn quantum theory of 
feedback \cite{wis1,wis2} using the input-output theory 
developed by Gardiner and Collett \cite{gar0,gar1}. 
The input-output formalism is essentially an 
Heisenberg approach for the whole system (cavity and
vacuum bath), in which the environment dynamics is described by the white-noise 
input operator $dB(t)$ satisfing the Ito rules \cite{gar0,gar1,gar2,gg1}:
\begin{equation}
    \begin{array}{c}
    dB(t)^{2}=dB^{\dagger}(t)^{2}=0, \; \; dB^{\dagger}(t)dB(t) = 0, \\ \\
    dB(t)dB^{\dagger}(t) = dt,
    \end{array}
\label{ito}
\end{equation}
and the following commutation' s relations:
\begin{equation}
    \begin{array}{c}
    \big[ \, dB(t), dB(t') \, \big] = 0 , \\ \\
    \big[ \, dB(t), dB^{\dagger}(t') \, \big] = \delta(t-t') dt dt'. 
    \end{array}
\label{com1}
\end{equation}

In the absence
of any feedback loop, the evolution of a generic cavity mode operator 
$\hat{O}(t)$ in the interaction picture is described
by the quantum Langevin equation \cite{gar0,gar1}:
\begin{eqnarray}
d \hat{O}(t)  &=&  \frac{ \gamma}{2} 
\big( \: 2 a^{\dagger}(t) \hat{O}(t) a(t) - a^{\dagger}(t) a(t) \hat{O}(t) 
- \hat{O}(t) 
a^{\dagger}(t)a(t) \: \big) \, dt \nonumber \\
& & - \: \sqrt{ \gamma} \, \big[\hat{O}(t), a^{\dagger}(t) \big]\, 
dB(t) \: + \: \sqrt{ \gamma} \, \big[\hat{O}(t), a(t)
\big]\, dB^{\dagger}(t),
\label{langevin}
\end{eqnarray}
where $\gamma$ is the cavity damping rate. 

Eq. (\ref{langevin}) can be solved explicitely
in terms of the evolution operator 
$U(t,t_{0})$
\begin{equation}
    \hat{O}(t) =  U^{\dagger}(t, 0) \hat{O} \, U(t, 0),
\label{langevinsol}
\end{equation}
which in the absence of feedback takes the following form \cite{gar2}
\begin{equation}
    U(t, t_{0}) = \, \stackrel{ \longleftarrow} 
    \exp \Big( \, \sqrt{ \gamma} \int_{t_{0}}^{t} dB^{\dagger}(t') a \; - \;
    \sqrt{ \gamma} \int_{t_{0}}^{t} dB(t') a^{\dagger} \Big),
\label{u1}
\end{equation}
where $a$ and $a^{\dagger}$ are in the Schr\"{o}dinger representation and
$\stackrel{ \longleftarrow} \exp $ denotes the time-ordered exponential. 
The evolution operator
$U(t, t_{0})$ describes also the evolution in the Schr\"{o}dinger 
representation, \cite{gar2,gg1},
\begin{equation}
| \Psi(t) \rangle_{T} = U(t,0) \;\Big\{ |\psi_{A}\rangle 
\otimes | 0 \rangle \Big\},
\label{psi}
\end{equation}
where the vector $| \Psi(t) \rangle_{T} $ obeys the following 
stochastic equation 
\begin{equation}
   d | \Psi(t) \rangle_{T}  = \Big\{ \sqrt{ \gamma} \, a \; 
   dB^{\dagger}(t) - \sqrt{ \gamma} \, a^{\dagger} \; dB(t)  -
                              \frac{\gamma}{2} \, a^{\dagger}a \,dt \Big\}	
                              | \Psi(t) \rangle_{T}.
\label{sscrhodinger}
\end{equation}
Using the commutation'~s rules (\ref{com1}), it is easy to prove that 
the Heisenberg  evolution (\ref{langevinsol}) 
satisfies the usual requirement that the input noise $dB(t)$ has to 
commute  
with every cavity operator evaluated at preceding times $t' < t$
\begin{equation}
    \big[\,dB(t),\hat{O}(t')\, \big] = \big[\,dB^{\dagger}(t),
    \hat{O}(t')\, \big] = 0 \hspace{.3in} \mbox{for} 
    \hspace{.2in} t \geq t'.
\label{com2}
\end{equation}

Equation (\ref{langevin}) can be used to get the time evolution of 
a generic matrix element of $\hat{O}(t)$ between two 
state vectors of the whole system of the form 
\mbox{$| \psi_{A} \rangle \otimes | 0 \rangle$} 
and \mbox{$| \psi_{B} \rangle \otimes | 0 \rangle$}, in which the environment is
left in the vacuum state,
\begin{equation}
   d \langle \hat{O}(t) \rangle_{\mbox{{\tiny  AB}}}  
   = \frac{ \gamma}{2} \left\{ \, 2  \langle a^{\dagger}(t)
   \hat{O}(t) a(t)  \rangle_{\mbox{{\tiny  AB}}} -  \langle
   a^{\dagger}(t) a(t) \hat{O}(t) \rangle_{\mbox{{\tiny  AB}}} 
   -  \langle \hat{O}(t)  a^{\dagger}(t)a(t) 
   \rangle_{\mbox{{\tiny  AB}}} \, \right\} \, dt,
\label{langevinmediata}
\end{equation}
where 
\begin{equation}
\langle \hat{O}(t) \rangle_{\mbox{{\tiny  AB}}} 
\equiv \Big\{ \langle 0 | \otimes \langle \psi_{A} | \Big\} \,\hat{O}(t) \, 
\Big\{ |\psi_{B}\rangle \otimes | 0 \rangle \Big\}.
\label{fuoridiagonale}
\end{equation}

Let us now introduce the feedback loop associated to the homodyne measurement 
of the quadrature $X_{\varphi}$. Differently from Ref.~\cite{wis3},
we assume the possibility of a non-unit homodyne detection efficiency 
$\eta \leq 1$. The application of a feedback loop 
is equivalent to add a feedback Hamiltonian 
$H_{fb}(t)$ 
\cite{wis1,wis2,qnd}, so that the correction to the Heisenberg evolution 
of Eq.~(\ref{langevin}) takes the form
\begin{equation}
    d \hat{O}_{{\em fb}}(t) = \frac{i \sqrt{\gamma} \, d Y_{\varphi}(t - 
    \tau)}{\eta} \, 
    \big[ F(t) , \hat{O}(t) \big]
\label{fbc}
\end{equation}
where $F(t)$ is the observable of the cavity mode through which the feedback 
acts on the system 
and $Y_{\varphi}(t)$ is the output field operator associated to
the homodyne measurement. In the quantum trajectory approach of 
\cite{wis1,wis2}, the fed-back homodyne photocurrent $Y_{\varphi}(t)$ 
is a classical quantity, but in the quantum Langevin approach it must 
be an operator with its quantum fluctuations. However,
one can adopt the general theory 
of homodyne measurements of Ref.~\cite{homo} and write 
the photocurrent operator in an analogous way
\begin{equation}
    d Y_{\varphi}(t) = 2 \, \sqrt{\gamma} \eta \, X_{\varphi}(t) \, 
    dt  \: + \sqrt{\eta}\: d \, \Xi_{\varphi}(t).
\label{dl}  
\end{equation} 
where $d\, \Xi_{\varphi}(t)$ describes the ``noisy'' part of the output
photocurrent operator. The only delicate point in the derivation of the
quantum theory of feedback of \cite{wis1,wis2} in the imperfect 
detection case using the input-output theory is just the exact 
determination of this noisy operator $d\, \Xi_{\varphi}(t)$. Since 
$Y_{\varphi}(t)$ is an output operator, it is quite natural to
consider Eq.~(\ref{dl}) as an input-output relation 
\cite{gar0,gar1,gg1}, so that the noisy term would simply be the input field
$d\, \Xi_{\varphi}(t) = dB(t) \, e^{- i \varphi} + dB^{\dagger}(t) 
\, e^{ i \varphi} $. However this interpretation of Eq.~(\ref{dl}) is 
correct only in the perfect detection case $\eta=1$, because only in 
this case the output of the detection apparatus coincides with the 
cavity output and the quantum fluctuations of the vacuum bath
are transferred unaltered by the detector. In the presence of 
imperfect detection, the output photocurrent may be non trivially 
related with the input noise $dB(t)$ and in general one has to 
describe the noisy operator $d\, \Xi_{\varphi}(t)$ in terms of a 
{\it new} noise $dB_{f}(t)$, which we shall call ``feedback''noise.
Therefore one has to write 
\begin{equation}
d\, \Xi_{\varphi}(t) = dB_{f}(t) \, e^{- i \varphi} + dB_{f}^{\dagger}(t) 
\, e^{ i \varphi} ,
\label{dXi}
\end{equation}
where the feedback noise $dB_{f}(t)$ satisfies the same properties 
(\ref{ito}) and (\ref{com1}) of the input noise; moreover this feedback
noise is correlated with the input noise $dB(t)$  and this 
correlation is determined just by the detection 
efficiency $\eta$, since one has
\begin{equation}
    \begin{array}{c}
    \big[ \, dB(t), dB_{f}(t') \, \big] = 0 , \\ \\
    \big[ \, dB(t), dB_{f}^{\dagger}(t') \, \big] = \sqrt{\eta}
    \delta(t-t') dt dt'. 
    \end{array}
\label{correta}
\end{equation}
It is immediate to see that in the perfect detection case
$\eta=1$, one can identify the feedback noise with the input noise
$dB_{f}(t)= dB(t)$, while in the opposite case $\eta=0$ the two
noises are uncorrelated, as it can be easily expected since in 
this case the fed-back noise has nothing to do with the vacuum input 
noise. 

In the feedback correction (\ref{fbc}) of 
the Heisenberg evolution, $\tau$ is the 
delay time associated to the feedback loop
and since it is a non-negative quantity, it ensures that the output
operator $Y_{\varphi}(t-\tau)$ commutes with all system operators 
evaluated at time $t$. In particular $Y_{\varphi}(t-\tau)$
commutes with $F(t)$ and so there is no ambiguity in the definition of 
$d \hat{O}_{{\em fb}}(t)$.
As it has been stressed in Ref. \cite{wis1,wis2,qnd}, one must be careful in 
using (\ref{fbc}); the
feedback process is physically added to the evolution of the system of 
interest, so its stochastic
differential contribution has to be introduced as limit of a real process. 
This implies that (\ref{fbc}) 
has to be considered in the Stratonovich sense. 
Therefore it is convenient to rewrite it in the Ito form and then add it
to Eq. (\ref{langevin}), so that 
the resulting equation for $\hat{O}(t) $ becomes
\begin{equation}
 \begin{array}{c}
   d \hat{O}(t)  =  \frac{ \gamma}{2} \big( \: 2 a^{\dagger}(t) \hat{O}(t)
    a(t) - a^{\dagger}(t) a(t) \hat{O}(t) - \hat{O}(t) 
          a^{\dagger}(t)a(t) \: \big) \, dt \\  \\
   - \: \sqrt{ \gamma} \, \big[\hat{O}(t), a^{\dagger}(t) \big]\, 
   dB(t) \: + \: \sqrt{ \gamma} \, \big[\hat{O}(t), a(t)
          \big]\, dB^{\dagger}(t) \\ \\
   + i \, \sqrt{ \gamma} \: \big( \: \sqrt{ \gamma} a^{\dagger}(t - \tau) 
   \, e^{i \varphi} \, dt +
          \frac{dB_{f}^{\dagger}(t -\tau)}{\sqrt{\eta}}
            e^{i \varphi} \: \big) \; \big[ F(t), 
          \hat{O}(t)  \big] \, \Theta(t - \tau) \\  \\
   + i \, \sqrt{ \gamma} \; \big[ F(t), \hat{O}(t)  \big] \; \big( \: 
   \sqrt{ \gamma} a(t - \tau) \, e^{ - i \varphi} \,
          dt + \frac{dB_{f}(t - \tau)}{\sqrt{\eta}}  e^{ - i \varphi} \: \big) \, 
          \Theta(t - \tau) \\ \\
   - \frac{\gamma}{2 \eta} \, \Theta(t - \tau) \big[ F(t), 
   \big[ F(t), \hat{O}(t) \big] \big] dt . 
 \end{array}
\label{l1}
\end{equation}
This is the quantum Ito stochastic equation describing the time
evolution in the presence of 
feedback and non-unit detection efficiency and it coincides
with the quantum Ito equation derived in Ref.~\cite{wis3} in the case
of perfect homodyne detection $\eta =1$.
We have also explicitely inserted the step function $ \Theta(t- \tau) $
with respect to Ref.~\cite{wis3} to stress the
impossibility for the feedback to act on the system before the delay time $\tau$ 
has elapsed since the initial condition. 
This means that the evolution of $\hat{O}(t)$  for $ 0\leq t\leq \tau$ 
coincides with that in absence of feedback, described by Eq.~(\ref{langevin}).
Moreover we observe that, since the equation for $d \, \hat{O}(t) $ 
contains only stochastic terms evaluated for times $t' \leq t$ 
(precisely $t$ and $t-\tau$), it is possible to conclude that
the commutation relations (\ref{com2}) are valid also in the presence 
of feedback and, more in general, that Eq.~(\ref{l1})
preserves the canonical commutation rules for $a(t)$ and $a^{\dagger}(t)$.

\subsection{Zero delay time limit}

Up to now, the explicit applications of the quantum theory of 
feedback of Wiseman and Milburn have considered the 
zero delay time case $\tau=0$ only, when one has 
a tractable Markovian equation. Whenever one considers a non-zero delay 
the problem becomes non-Markovian and difficult to solve.

The feedback master equation for homodyne-mediated feedback in the 
zero delay limit has been first derived in its general form
using quantum trajectory theory 
\cite{carm} in Ref.~\cite{wis1,wis2}. In the case
of perfect homodyne detection $\eta=1$,
the same homodyne-mediated
feedback equation has been rederived using input-output theory by Wiseman 
in \cite{wis3} and, 
in its {\em linear} stochastic form, by Goetsch {\it et al}. in 
Ref.~\cite{GTV}. However, the connection between this 
linear stochastic approach and the input-output theory was not 
made explicit there. In reviewing the zero delay time case we shall 
clarify here the connections between the different approaches and we
show in particular that the linear stochastic Schr\"odinger equation 
approach of Ref.~\cite{GTV} is equivalent to the input-output result 
of Eq.~(\ref{l1}) (in the case $\eta=1$) 
in the same way as Eq.~(\ref{sscrhodinger}) is 
equivalent to the quantum Langevin equation Eq.~(\ref{langevin}).
 
The starting point of the analysis of Ref. \cite{GTV} is the evolution 
equation for the state vector $ | \Psi(t)
\rangle_{T}$  of the whole system (cavity and vacuum bath). 
In the no feedback case, this equation is obviously
equal to Eq. (\ref{sscrhodinger}); the feedback loop is then 
introduced using the same Hamiltonian modification 
of (\ref{fbc}) in the case $\eta=1$. 
However, since in Ref.~\cite{GTV} the zero delay time limit
is considered from the beginning, one has to be careful with operator 
ordering, because in this circumstance 
one is not guaranteed that $d Y_{\varphi}(t)$  commutes with the 
cavity mode operator $F(t)$ at the
same time. In Ref.~\cite{GTV} the question is solved
by imposing that ``the feedback acts later'', i.e.,
that $ | \Psi(t + dt) \rangle_{T}$ is obtained from $|\Psi(t)
\rangle_{T}$ by means of $U(t+dt,t)$ 
(the evolution operator in the absence of feedback loop) first 
and by the Hamiltonian feedback correction later (see Eq. (2.11) 
of \cite{GTV}). It is clear that in this way the equivalence with the
input-output approach of Ref. \cite{wis3} it is not so evident. 
However the equivalence can be proved by considering the following 
evolution operator
\begin{equation}
   \begin{array}{r}
    U_{F}(t, t_{0}) = \, \stackrel{ \longleftarrow} \exp \Big\{ \int_{t_{0}}^{t}
     \Big( \sqrt{ \gamma} 
 dB^{\dagger}(t') a \; - \; \sqrt{ \gamma} dB(t') a^{\dagger} - i 
 \sqrt{ \gamma} F e^{-i\varphi} dB(t')\nonumber \\ \
     -i \sqrt{ \gamma} F e^{i \varphi} dB^{\dagger}(t') - i \frac{\gamma}
     {2} a^{\dagger} F e^{i
    \varphi} d t' - i \frac{\gamma}{2} F a e^{-i \varphi} d t' \Big) \Big\};
   \end{array}
\label{u2}
\end{equation}
where $a$, $a^{\dagger}$ and $F$ are in Schr\"{o}dinger representation.
Using the Ito rules (\ref{ito}) to evaluate the differential 
$d \hat{O}(t)$, it is possible to check that $U_{F}(t, t_{0})$ is just
the evolution operator determining the formal solution of Eq.~(\ref{l1})
in the $\tau=0$ and $\eta=1$ limit, according to the usual rule
\begin{equation}
\hat{O}(t) = U_{F}^{\dagger}(t, 0) \, \hat{O} \, U_{F}(t, 0)\;.
\label{evdio}
\end{equation}
As it can be easily expected, the zero delay evolution operator $U_{F}(t, t_{0})$ 
reduces to the no-feedback one $U(t,t_{0})$ of Eq.~(\ref{u1}) 
when $F=0$. This suggests that an equivalent Schr\"odinger 
representation could be obtained also in the case of feedback with
zero delay time,
starting from an equation analogous to Eq.~(\ref{psi}).
In fact if we apply  $U_{F}(t, 0)$ to the initial state 
of the total system and we use 
again the Ito rules (\ref{ito}), one gets the following linear 
stochastic Schr\"odinger equation,
\begin{eqnarray}
   d | \Psi(t) \rangle_{T}  &=& \Big\{ \sqrt{ \gamma} \, a \; dB^{\dagger}(t) 
   - \sqrt{ \gamma} \, a^{\dagger} \; dB(t)  -
                              \frac{\gamma}{2} \, a^{\dagger}a \,dt \nonumber\\ 
                              \nonumber \\
  & & -i \sqrt{\gamma} \, F \; d \, \Xi_{\varphi}(t)  -i \gamma F \, 
  a e^{- i \varphi} dt - \frac{\gamma}{2} \, F^{2} \, dt
                              \Big\}	| \Psi(t) \rangle_{T}
\label{GTV1}
\end{eqnarray}
coinciding with the equation obtained in \cite{GTV}. This shows that the 
approach of Ref. \cite{GTV} and that of Ref. \cite{wis3} are respectively 
the Schr\"{o}dinger and Heisenberg view of
the same theory, with $U_{F}(t, t_{0})$ the unitary operator mediating the 
transition from one to the other.

In Ref.~\cite{GTV}, by adopting an appropriate 
representation basis for the vacuum modes (see for example \cite{gg1}), 
Eq.~(\ref{GTV1}) 
was then reduced to a linear stochastic equation for the cavity mode only 
which was solved numerically. In Appendix A we shall reconsider this
linear stochastic equation for the cavity mode and we shall see
how it is possible to solve it analytically by adopting the 
integration method described in \cite{jac}.

Determining an evolution operator analogous to $U_{F}(t, t_{0})$ for 
the $\tau \neq 0$ case is much more difficult and we shall not 
consider this strategy to study the non-zero delay problem.
Instead we shall adopt the input-output formalism which has yielded 
Eq.~(\ref{l1}). To be more specific, we shall always consider generic 
matrix elements as those of Eq.~(\ref{fuoridiagonale}), whose evolution
equation can be easily derived from Eq.~(\ref{l1}):
\begin{equation}
 \begin{array}{c}
  \frac{d}{dt}\langle \hat{O}(t) \rangle_{\mbox{{\tiny  AB}}} = 
  \frac{ \gamma}{2} \left\{ \: 2  
       \langle a^{\dagger}(t) \hat{O}(t) a(t) \rangle_{\mbox{{\tiny  
       AB}}} -
       \langle a^{\dagger}(t) a(t) \hat{O}(t) \rangle_{\mbox{{\tiny  
       AB}}} - 
       \langle \hat{O}(t)  a^{\dagger}(t) a(t)  \rangle_{\mbox{{\tiny  
       AB}}} \: \right\} \\ \\
   + i \, \gamma  \left\{ \langle a^{\dagger}(t - \tau)  \big[ F(t), 
       \hat{O}(t)  \big] \rangle_{\mbox{{\tiny  AB}}} \, 
       e^{i \varphi} \, + \,
       \langle \big[ F(t), \hat{O}(t)  \big] a(t - \tau) 
       \rangle_{\mbox{{\tiny  AB}}}
       \, e^{ - i \varphi} \right\} \Theta(t - \tau)  \\ \\
   - \frac{\gamma}{2 \eta} \, \langle \big[ F(t), \big[ F(t), 
       \hat{O}(t) \big] \big] \rangle_{\mbox{{\tiny  AB}}} \;.
       \Theta(t -\tau)  .
 \end{array}  
\label{l1m}
\end{equation}

\section{Feedback dynamics in the presence of a non-zero delay}

The dynamics in the presence of a feedback loop with a non-zero delay 
time has never been completely solved because of its intrinsic non-Markovian 
nature. In this paper we shall analyze the effects of a non-zero 
feedback delay by considering a specific example for the ``feedback 
operator'' $F(t)$
\begin{equation}
 F(t) = g \, X_{\theta}(t) = \frac{g}{2} ( \: a(t) \, 
 e^{ - i \theta} \,+ \, a^{ \dagger}(t) \, e^{  i \theta} \: ),
        \label{qfb}
\end{equation}
where the constant $g$ represents the gain of the feedback process and
 $\theta$ is an experimentally controllable phase. 
The particular choice (\ref{qfb}) of $F$ means
that the feedback loop adds a driving term to the mode dynamics, which could
be achieved, e.~g.~, by using an electro-optic device with variable
transmittivity driven by the homodyne photocurrent. The 
homodyne-mediated feedback model with the choice (\ref{qfb}) for 
$F(t)$ has been completely solved in Ref.~\cite{GTV} in the Markovian 
limit of zero delay time and therefore the comparison with the 
results of Ref.~\cite{GTV} will be very instructive.
As it is shown in Ref.~\cite{GTV}, the main virtue of the 
homodyne-mediated feedback is its capability of slowing 
down the decoherence associated with cavity damping provided that the 
feedback parameters $g$ and $\theta$ are appropriately chosen. Here we 
shall see that the decoherence inhibition caused by the feedback takes 
place also in the presence of a non-zero feedback delay time; in 
particular, decoherence is 
appreciably slowed down even for delay times not much smaller than the 
decoherence time itself. 

First of all we shall show the exact time evolution for the marginal 
probability distribution $P(x_{\varphi},t)$ 
of the quadrature component $X_{\varphi}(t)$:
this will result in a quite simple expression which can be easily 
analysed. Then we shall give the complete solution of the system 
dynamics in terms of the symmetrically ordered characteristic 
function.

\subsection{The marginal probability distribution}

A good, even if not complete, description of the state of the cavity mode
is given by the marginal probability distribution $P(x_{\varphi},t)$
of the measured quadrature component $X_{\varphi}(t)$.
We shall consider the following class of initial states 
for the whole system
\begin{equation}
  \rho_{T}(0) = 
  \sum_{\mbox{{\tiny  $\alpha,\beta$}}} N_{\mbox{{\tiny  $\alpha,\beta$}}}
  | \alpha \rangle \langle \beta | \otimes |0 \rangle \langle 0 |,
\label{condin}
\end{equation}
i.e. a linear superposition of coherent states for the cavity 
mode and the vacuum state for the electromagnetic bath. We shall 
focus on the evaluation of the moments
\begin{equation}
	\langle  X_{\varphi}^{N}(t)   \rangle_{\mbox{{\tiny  $\beta 
	\alpha$}}}\equiv \Big\{ \langle 0 | \otimes \langle \beta | \Big\} \, 
	X_{\varphi}^{N}(t) \, 
    \Big\{ |\alpha \rangle \otimes | 0 \rangle \Big\}
	\label{neq0}
\end{equation}
($N$ is an integer). $P(x_{\varphi},0)$ is a superposition of 
Gaussian functions and, thanks to the choice of Eq. (\ref{qfb}) for the 
feedback operator, the evolution in the presence of feedback remains 
linear, so that $P(x_{\varphi},t)$ will maintain 
its initial Gaussian behaviour. This fact will be explicitely verified 
at the end of the section. 

We now proceed step by step: first we explicitely determine 
the evolution of the first and second order moment
and then we derive a recursive 
relation between the moments clearly showing the Gaussian nature of
the corresponding probability distribution.

\begin{figure}
\centerline{\epsfig{figure=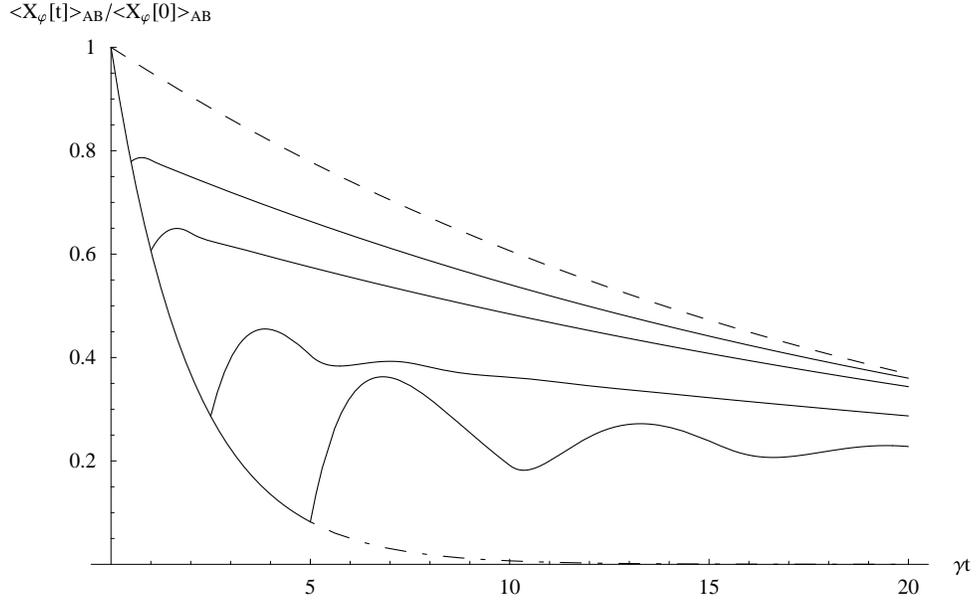,width=13cm}}
\vspace{0.2cm}
\caption{Normalized mean value of the measured quadrature
$\chi(t)$ (see Eq.~(\ref{soluzione}))
 for $g\sin(\theta-\varphi)=0$ (without
feedback, dot-dashed line), $g\sin(\theta-\varphi)=0.45$ and some values of 
the delay time: from bottom to top: $\gamma
\tau=5$, $\gamma \tau=2.5$, $\gamma \tau=1$, $\gamma \tau= 0.5$ (full lines) 
and $\gamma \tau=0$ (dashed line).}
\label{Xmedio}
\end{figure}

For $N=1$ Eq. (\ref{neq0}) becomes the matrix 
element of the measured quadrature $X_{\varphi}(t)$, and Eqs.~(\ref{l1m}) 
and (\ref{qfb}) yield the following differential equation for its 
evolution
\begin{equation}
  \frac{d}{dt}\langle X_{\varphi}(t) \rangle _{\mbox{{\tiny 
   AB}}} = - \frac{ \gamma}{2} \langle X_{\varphi}(t) 
  \rangle _{\mbox{{\tiny AB}}} + \gamma \, g \sin(\theta -
  \varphi) \langle X_{\varphi}(t - \tau) \rangle _{\mbox{{\tiny 
  AB}}} \Theta(t - \tau),
\label{qfi}
\end{equation}
which can be integrated for $t\geq 0$ using Laplace transforms as it is
discussed in Appendix B:
\begin{equation}
 \langle X_{\varphi}(t) \rangle _{\mbox{{\tiny AB}}} = 
 \langle X_{\varphi}(0) \rangle _{\mbox{{\tiny AB}}} \chi(t)
\label{soluzione}
\end{equation}
with 
\begin{equation}
	\chi(t)=\sum_{n=0}^{[t/ \tau]} \frac{ \big( g \sin( \theta -
\varphi) \big)^{n}}{n!}\, e^{ -\frac{\gamma}{2} (t - n \tau)} \, 
\big( \gamma \, (t - n \tau)\big)^{n},
	\label{neq1}
\end{equation}
where $[x]$ indicates the integer part of the real number $x$. 
In the case of the initial condition (\ref{condin}),
one has simply to consider $|\psi_{A}\rangle \equiv | \beta \rangle$ and $| 
\psi_{B}\rangle = | \alpha \rangle $ in Eq. (\ref{soluzione}), 
even though it is clear that this solution is valid for any 
choice of the initial state of the cavity. The function $\chi(t)$ of 
Eq.~(\ref{neq1}) will often appear in the complete analytical 
solution of the problem described in the following and it is therefore 
useful to describe its behavior in the various limits.
The solution (\ref{soluzione}) has the correct behavior both in the 
$g \rightarrow 0$ limit, where one has the simple exponential decay
$\chi(t)= e^{-\gamma t/2}$, and in the
$ \tau \rightarrow 0$ limit, yielding
\begin{equation}
 \chi(t) =
  \exp \Big\{- \frac{\gamma}{2} \, \big( 1 - 2 g \sin(\theta -\varphi) \big )
 \, t \Big\},
\label{soluzionenodelay}
\end{equation}
which is the same solution which can be derived from the exact treatment of 
Ref.~\cite{GTV}. It is interesting to consider the limit of small delay, 
$\gamma \tau \ll 
1$, because this condition can be easily verified using good optical 
cavities and common electro-optical feedback loops. 
From the exact solution (\ref{neq1}) one gets
\begin{equation}
\chi(t) = \Big\{ 1 - \frac{g \sin(\theta-\varphi)}{2}\,\Big(2 -
 \gamma t  (1-2g\sin(\theta-\varphi)) \Big) \gamma \tau \Big\}\,
e^{-(1 -2 g\sin(\theta-\varphi))\frac{\gamma}{2}t}.
\label{chi2}
\end{equation} 
We have plotted the solution (\ref{neq1})
in Fig.~\ref{Xmedio}, in which there is also a comparison 
with the no feedback case and with the Markovian feedback case of the 
zero delay time limit. This plot makes evident how the major part of the 
difference between the zero delay and the delayed cases comes
from the retardation caused by the
presence of $\Theta(t - \tau)$: for simplicity, we shall refer 
in the following to this effect as the ``step effect''. 

For the determination of the second order moment (Eq.~(\ref{neq0}) with 
$N=2$) it is convenient to study the correlation function 
$C(t,t')=\langle X_{\varphi}(t) X_{\varphi}(t')\rangle_{\mbox{{\tiny $\beta 
\alpha$}}}$ for every positive values of $t$ and $t'$. 
We first focus on the dependence on $t'$ and differentiate $C(t,t')$ by 
$t'$ using Eq.~(\ref{l1}), so to get the following differential 
equation:
\begin{eqnarray}
 \label{neq2}
 d \; C(t,t') &=& - \frac{ \gamma}{2} C(t,t') \, dt'
   + \gamma \, g \sin(\theta - \varphi) C(t,t'-\tau) \Theta(t' - 
   \tau) \, dt' \\ \nonumber \\
 & & -\frac{\sqrt{\gamma}}{2}e^{i\varphi} \langle \big[ 
 X_{\varphi}(t),dB^{\dagger}(t') \big] \rangle_{\mbox{{\tiny $\beta \alpha$}}}
     +\sqrt{\frac{\gamma}{4\eta}} g \sin(\theta - \varphi) e^{i\varphi} 
     \langle \big[X_{\varphi}(t), dB_{f}^{\dagger}(t'-\tau) \big]\rangle_{\mbox{{\tiny $\beta 
     \alpha$}}} \Theta(t'-\tau). \nonumber
\end{eqnarray}
The corresponding initial condition can be obtained from Eq.~(\ref{soluzione}) 
and is given by 
$C(t,0)=\langle X_{\varphi}(0) X_{\varphi}(0)\rangle_{\mbox{{\tiny $\beta 
\alpha$}}} \, \chi(t)$. Now Eq.~(\ref{neq2})
can be explicitely integrated once that the explicit $t'$ dependence of 
the commutators between the noises $dB^{\dagger}(t')$ and 
$dB_{f}^{\dagger}(t'-\tau)$ 
and the operator $X_{\varphi}(t)$ is known. For the determination of 
these commutators, let us define first of all $f(t,t')=\big[X_{\varphi}(t), 
 B^{\dagger}(t')\big]$, in which $B^{\dagger}(t')$ is the sum of all the Ito 
 increments of the input noise from the initial time up to $t'$. Now we can 
proceed in an analogous way as we have done to 
 write Eq.(\ref{neq2}) and we differentiate in $t$ by
keeping $t'$ constant. 
Using the commutation rules of Eq.~(\ref{com1}) we get
 \begin{eqnarray}
 \frac{\partial}{\partial t} f(t,t') &=& - \frac{ \gamma}{2} f(t,t') 
   + \gamma \, g \sin(\theta - \varphi) f(t-\tau,t') \Theta(t - 
   \tau) \nonumber \\  \label{neq3} \\
 & & -\frac{\sqrt{\gamma}}{2}e^{-i\varphi}\Big\{ \Theta(t'-t)
     - g \sin(\theta - \varphi) \Theta(t-\tau) \Theta(t'-t+\tau) 
     \Big\}.\nonumber
\end{eqnarray}
which, except for the presence of last term which is a known 
function of time, is an equation 
similar to Eq.~(\ref{qfi}) and therefore can be solved using Laplace 
transforms and the initial condition  
$f(0,t')=0$ implied by Eq.~(\ref{com2}). 
The commutation rules between $X_{\varphi}(t)$ and
the Ito increment $dB^{\dagger}(t')$ can now be obtained by
simply differentiating this 
solution with
respect to $t'$ and the final result is
\begin{equation}
   \big[X_{\varphi}(t), dB^{\dagger}(t')\big]={\cal F}(t,t') \, dt' 
   \label{neq5}
\end{equation}
where
\begin{equation}
	{\cal F}(t,t')=-\frac{\sqrt{\gamma}}{2}e^{-i\varphi}\Big\{ 
	\Theta(t-t') \chi(t-t') - g \sin(\theta- \varphi) \, 
	\Theta(t-t'-\tau)\; \chi(t-t'-\tau) \Big\}\equiv{\cal F}(t-t').
	\label{neq7}
\end{equation}
The same procedure can be adopted to determine the commutator
$\big[X_{\varphi}(t), dB_{f}^{\dagger}(t')\big]$ involving the 
feedback noise and one finally gets
\begin{equation}
   \big[X_{\varphi}(t), dB_{f}^{\dagger}(t')\big]={\cal F}_{f}(t,t') \, dt' 
   \label{neq5bis}
\end{equation}
where
\begin{equation}
	{\cal F}_{f}(t,t')=-\frac{\sqrt{\gamma}}{2}e^{-i\varphi}\Big\{ 
	\sqrt{\eta}\Theta(t-t') \chi(t-t') - \frac{g \sin(\theta- 
	\varphi)}{\sqrt{\eta}} \, 
	\Theta(t-t'-\tau)\; \chi(t-t'-\tau) \Big\}\equiv{\cal F}_{f}(t-t').
	\label{neq7bis}
\end{equation}

Proceeding as before it is possible to compute also the 
commutation rules between the noise operators $dB(t')$, 
$dB^{\dagger}(t')$, $dB_{f}(t')$, 
$dB_{f}^{\dagger}(t')$ and  $a(t)$, $a^{\dagger}(t')$ which we explicitely
report here because they 
will be useful in the following
\begin{equation}
	\begin{array}{ll}
		 \big[a(t), dB(t')\big]={\cal R}(t,t') \, dt' \hspace{.2in} &\big[a(t), 
		 dB^{\dagger}(t')\big]=-{\cal R}_{1}^{*}(t,t') \, dt'\\ \\ 
		 \big[a^{\dagger}(t), dB(t')\big]={\cal R}_{1}(t,t') \, dt' \hspace{.2in} & 
		  \big[a^{\dagger}(t), dB^{\dagger}(t')\big]=-{\cal R}^{*}(t,t') \, 
		  dt', \\ \\
		   \big[a(t), dB_{f}(t')\big]={\cal R}^{f}(t,t') \, dt' \hspace{.2in} &\big[a(t), 
		 dB_{f}^{\dagger}(t')\big]=-{\cal R}_{1}^{f\,*}(t,t') \, dt'\\ \\ 
		 \big[a^{\dagger}(t), dB_{f}(t')\big]={\cal R}_{1}^{f}(t,t') \, dt' \hspace{.2in} & 
		  \big[a^{\dagger}(t), dB_{f}^{\dagger}(t')\big]=-{\cal R}^{f\,*}(t,t') \, 
		  dt',
	\end{array}
	\label{neq5A}
\end{equation}
where 
\begin{eqnarray}
	{\cal R}(t,t') &=& i \frac{\sqrt{\gamma}}{2} \, g \, e^{i(\theta +\varphi)}\Big\{ 
	\Theta(t-t'-\tau) \Big[ e^{-\frac{\gamma}{2}(t-t'-\tau)} - 
	\frac{\chi(t-t') -  e^{-\frac{\gamma}{2}(t-t')}}{g 
	\sin(\theta-\varphi)}\Big] \label{neq7B} \\ \nonumber \\
	& & + \Theta(t-t'-2 \tau) \big[ \chi(t-t'-\tau) -  
	e^{-\frac{\gamma}{2}(t-t'-\tau)}\big] \Big\}\equiv{\cal R}(t-t'), \nonumber
\end{eqnarray}
\begin{equation}
	{\cal R}_{1}(t,t') = -2e^{-i \varphi}{\cal F}^{*}(t,t') - e^{-2 i 
	\varphi} {\cal R}(t,t')\equiv{\cal R}_{1}(t-t').  
	\label{neq7A}
\end{equation}
\begin{eqnarray}
	{\cal R}^{f}(t,t') &=& i \sqrt{\frac{\gamma}{4\eta}} \, g \, e^{i(\theta +\varphi)}\Big\{ 
	\Theta(t-t'-\tau) \Big[ e^{-\frac{\gamma}{2}(t-t'-\tau)} - 
	\frac{\eta \left(\chi(t-t') -  e^{-\gamma (t-t')/2}\right)}{g 
	\sin(\theta-\varphi)}\Big] \label{neq7Bbis} \\ \nonumber \\
	& & + \Theta(t-t'-2 \tau) \big[ \chi(t-t'-\tau) -  
	e^{-\gamma (t-t'-\tau)/2}\big] \Big\}\equiv{\cal R}^{f}(t-t'), \nonumber
\end{eqnarray}
\begin{equation}
	{\cal R}_{1}^{f}(t,t') = -2e^{-i \varphi}{\cal F}_{f}^{*}(t,t') - e^{-2 i 
	\varphi} {\cal R}^{f}(t,t')\equiv{\cal R}_{1}^{f}(t-t').  
	\label{neq7Abis}
\end{equation}
Note that all these commutators are c-number functions and this 
is essentially a consequence of the commutation 
rules between the noise operators given by Eqs.~(\ref{com1}) and (\ref{correta}).
At this point it is possible to compute the correlation function 
$C(t,t')$ replacing Eq.~(\ref{neq5}) in Eq.~(\ref{neq2}): we obtain 
 an integrable differential equation of the same form of 
Eq.~(\ref{neq3}), whose solution is
\begin{equation}
	C(t,t')=C(0,0) \; \chi(t) \chi(t') + \langle \beta|\alpha \rangle {\cal G}(t,t'),
	\label{neq8}
\end{equation}
where
\begin{eqnarray}
	{\cal G}(t,t')&=&\frac{\gamma}{4} \Big\{ 
	\int_{0}^{\min[t,t']}d t^{\prime \prime} \;\chi(t-t^{\prime 
	\prime})\chi(t'-t^{\prime \prime}) \nonumber \\ \nonumber\\
    & & -g \sin(\theta -\varphi)  \Theta(t'-\tau) 
    \int_{0}^{\min[t,t'-\tau]}d t^{\prime \prime} \;
    \chi(t-t^{\prime \prime}) \chi(t'-\tau-t^{\prime \prime}) 
    \nonumber \\ \label{neq9}\\
    & & -g \sin(\theta -\varphi)  \Theta(t-\tau) 
    \int_{0}^{\min[t-\tau,t']}d t^{\prime \prime}\;
	\chi(t-\tau-t^{\prime \prime}) \chi(t'-t^{\prime \prime}) \nonumber \\ \nonumber \\
	& & + \frac{g^{2} \sin^{2}(\theta-\varphi)}{\eta} \Theta(t-\tau) \Theta(t'-\tau) 
	\int_{0}^{\min[t-\tau,t'-\tau]}d t^{\prime \prime}\;
	\chi(t-\tau-t^{\prime \prime})\chi(t'-\tau-t^{\prime \prime}) \Big\}.
	\nonumber 
\end{eqnarray}
If we now set $t'=t$ in this expression, we get the second order moment 
(Eq.~(\ref{neq0}) with $N=2$); it is however more useful to 
consider the expression of the following ``variance''
 \begin{eqnarray}
  	\sigma^{(2)}(t)&=& 2 \, \Big\{ \;\frac{ \langle X_{\varphi}^{2}(t) \rangle_{\mbox{{\tiny 
    $\beta \alpha$}}}}{\langle \beta|\alpha \rangle} - \Big( \frac{ \langle X_{\varphi}(t) \rangle_{\mbox{{\tiny 
    $\beta \alpha$}}}}{\langle \beta|\alpha \rangle}\Big)^{2} \; 
    \Big\} \label{neq10} \\ \nonumber \\
    &=& \frac{1}{2} + \frac{\gamma}{2 \eta} g^{2} 
  	\sin^{2}(\theta -\varphi) \Theta(t-\tau) \int_{0}^{t-\tau}d t^{\prime 
  	\prime}\; \chi^{2}(t^{\prime \prime}).
 	\label{neq11}
 \end{eqnarray}
Let us consider again the physically interesting limit of small delay, 
$\gamma \tau \ll 1$, in which
the variance of Eq.~(\ref{neq11}) 
can be well approximated by the first order expansion in 
$\gamma \tau$, which is given by 
 \begin{eqnarray}
\lefteqn{\sigma^{(2)}(t) = \frac{1}{2} \left\{ 1 + \frac{g^{2}}{\eta} 
\sin^{2}(\theta-\varphi) \left( \frac{ 1
- e^{-\gamma (1 -2 g \sin(\theta-\varphi))t}}{1 - 2 g \sin(\theta-\varphi)} \right) \right\} } \label{sigma2} 
\\ \nonumber \\ 
&&-\frac{g^{2} \sin^{2}(\theta-\varphi)}{2\eta} \! \left\{ \!(1 + \gamma t \, g \sin(\theta-\varphi)) e^{-\gamma (1 -2
g \sin(\theta-\varphi))t} + g \sin(\theta-\varphi) \left( \frac{ 1
- e^{-\gamma (1 -2 g \sin(\theta-\varphi))t}}{1 - 2 g 
\sin(\theta-\varphi)} \right)\! \right\} \gamma \tau \;.  \nonumber
\end{eqnarray}
This expression, as well as Eq.~(\ref{chi2}), is valid for 
$t>\tau$ only, since for $t < \tau $ the feedback is not yet acting 
and the variance assumes its value in absence of feedback, 
$\sigma^{(2)}(t)= 1/2$. 

As concerns the higher order variances, it is convenient to
consider the following quantities
\begin{equation} 
  \langle \hat{X}_{\varphi}^{N}(t) \rangle_{\mbox{{\tiny $\beta \alpha$}}}=
  \langle \Big( \; X_{\varphi}(t) - \chi(t) \, 
  X_{\varphi}(0) \; \Big)^{N}\rangle_{\mbox{{\tiny $\beta \alpha$}}}
  \label{neq11B}
\end{equation}
and to proceed as in the previous case, that is, by considering the function 
  $\langle \hat{X}_{\varphi}^{N-1}(t) 
    \hat{X}_{\varphi}(t')\rangle_{\mbox{{\tiny $\beta \alpha$}}}$,
and differentiating it with respect to $t'$ by keeping $t$ constant. This gives a 
differential equation which can be formally integrated and setting 
then $t=t'$ it is easy to get the following recursive relation: 
  \begin{equation}
   	\langle \hat{X}_{\varphi}^{N}(t) \rangle_{\mbox{{\tiny $\beta 
   	\alpha$}}}= (N-1) {\cal G}(t,t) \langle \hat{X}_{\varphi}^{N-2}(t) 
   	\rangle_{\mbox{{\tiny $\beta \alpha$}}}.
   	\label{neq13}
   \end{equation}
This relation can be easily solved and it can be expressed in the 
following way
\begin{eqnarray}
 \langle \Big( \; X_{\varphi}(t) - \frac{\langle 
 X_{\varphi}(t) \rangle_{\mbox{{\tiny $\beta \alpha$}}}}{\langle 
 \beta|\alpha \rangle} \Big)^{N}\rangle_{\mbox{{\tiny $\beta 
 \alpha$}}}&=& \langle \Big( \; X_{\varphi}(t) - \frac{\alpha 
 e^{-i\varphi}+ \beta^{*}e^{i\varphi}}{2} 
\chi(t) \Big)^{N}\rangle_{\mbox{{\tiny $\beta 
 \alpha$}}} \label{neq14A} \\ \nonumber \\
&=&\langle \beta|\alpha \rangle \left\{
  \begin{array}{ll}
  	0 & \hspace{.2in} \mbox{for} \hspace{.2in}  N  \hspace{.2in} \mbox{odd}  \\ \\ 
  	\frac{(N-1)!!}{2^{N/2}} \; \big[ \sigma^{(2)}(t) \big]^{N/2} & \hspace{.2in} \mbox{for} 
  	\hspace{.2in} N  \hspace{.2in} \mbox{even}, 
  \end{array} \right.
	\label{neq14}
\end{eqnarray}
reproducing the results for the mean and the variance derived above 
for $N=1$ and $N=2$ respectively.

The moments of Eq.~(\ref{neq14}) satisfy the typical relation of a 
Gaussian process and this provides an independent check of the fact 
that the probability distribution $P(x_{\varphi},t)$, being a 
superposition of Gaussians at $t=0$, remains Gaussian at all times, as 
it must be, due to the linearity of the evolution equation.  
This probability distribution can be written in 
the following form
\begin{equation}
P(x_{\varphi},t) = \sum_{\mbox{{\tiny  $\alpha,\beta$}}} N_{\mbox{{\tiny  $\alpha,\beta$}}} \;
\frac{\langle \beta | \alpha \rangle}{\sqrt{\pi \sigma^{(2)}(t)}} \, 
\exp\left\{-\frac{\big[\,x_{\varphi}- \frac{\alpha e^{-i \varphi} + 
\beta^{*}e^{i \varphi}}{2} 
\chi(t)\,\big]^{2}} {\sigma^{(2)}(t)} \right\},
\label{solup}
\end{equation}
with $\chi(t)$ and $\sigma^{(2)}(t)$ given respectively by Eq.~(\ref{neq1}) 
and Eq.~(\ref{neq11}). If we set $\tau=0$ in these expressions, 
we obtain the exact 
solution of the ideal case of zero feedback delay, which has been 
derived in \cite{GTV}.

\begin{figure}
\centerline{\epsfig{figure=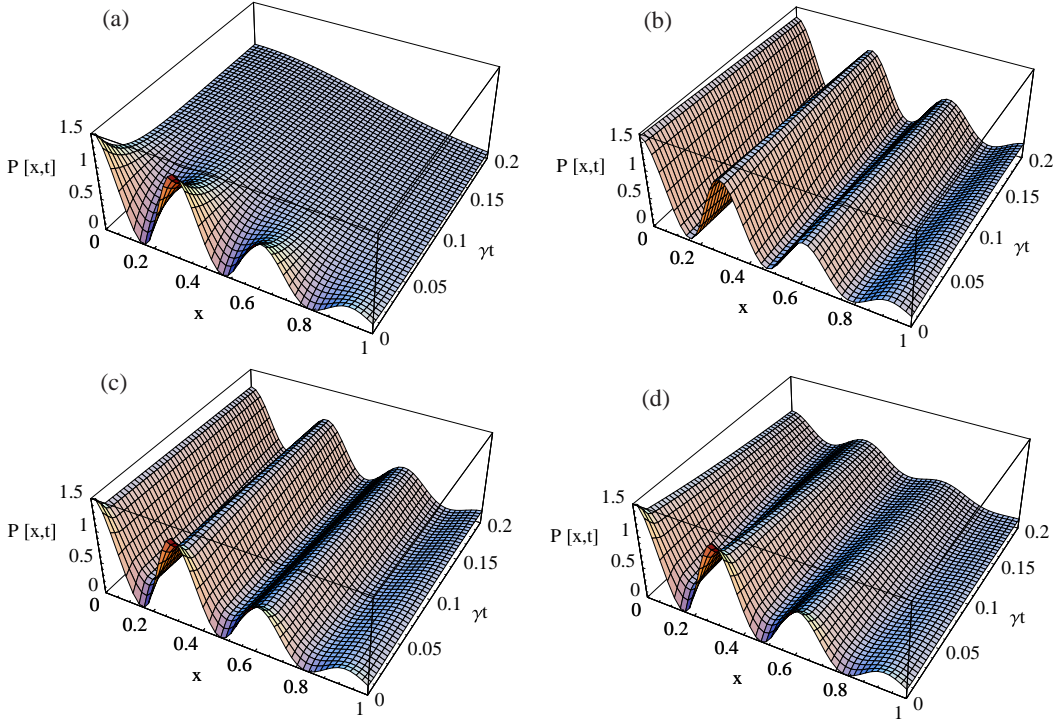,width=14cm}}
\vspace{0.2cm}
\caption{Time evolution of the marginal probability distribution 
$P(x,t)$ ($x=x_{\varphi}$) for the initial state (\ref{gatto1}) with 
$\alpha_{0}=i5$. (a) refers to the no feedback case
($g\sin(\theta-\varphi)=0$); (b) refers to the ideal limit
of zero feedback delay time and $\eta =1$, 
with $g\sin(\theta-\varphi)=1$; (c) refers to the case
with a non-zero feedback delay time ($\gamma
\tau=0.01$) $\eta =1$ and with $g\sin(\theta-\varphi)=1$; (d) refers to
the case $\gamma \tau=0.01$, $\eta =0.9$ and $g\sin(\theta-\varphi)=1$.}
\label{Pdix}
\end{figure}

\begin{figure}
\centerline{\epsfig{figure=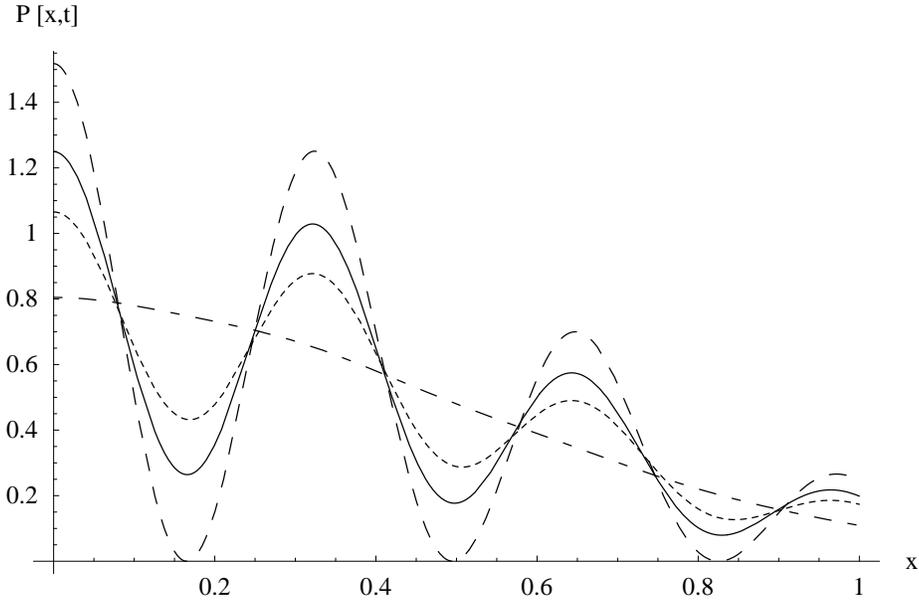,width=13cm}}
\vspace{0.2cm}
\caption{Comparison at $\gamma t= 0.1$ of the marginal probability 
distribution $P(x,t)$ between the cases of no feedback
(dot-dashed line), feedback with zero delay time 
($g\sin(\theta-\varphi)=1$, $\eta=1$, dashed line), feedback with non-zero
delay time $\gamma
\tau = 0.01$ and $\eta=1$ ($g\sin(\theta-\varphi)=1$, full line),
and feedback with non-zero delay time and imperfect detection
$\gamma \tau = 0.01$ and $\eta=0.9$ ($g\sin(\theta-\varphi)=1$, dotted line)}
\label{Pdixtfisso}
\end{figure}

It is instructive to apply the general result of Eq.~(\ref{solup})
to the case of an initial even Schr\"{o}dinger-cat state
\begin{equation}
| \Psi_{\em Cat} \rangle_{T} =  N \Big\{| \alpha_{0} \rangle \: + 
\: |- \alpha_{0}  \rangle  \Big\} \otimes
| 0
\rangle,
\label{gatto1}
\end{equation}
in which $|\pm \alpha_{0} \rangle $ are two coherent states of the cavity 
mode and $N=(1+e^{-2|\alpha_{0}|^{2}})^{-1/2}$ is the normalization constant, 
to see the effect of the 
non-zero delay on the decoherence process.
The marginal probability distribution $P(x_{\varphi},t)$ for this initial 
condition can be written as (see also Ref. \cite{GTV}) 
\begin{equation}
P(x_{\varphi},t)=N^{2} \{ p^{2}_{+}(x_{\varphi},t) + p^{2}_{-}(x_{\varphi},t) +2p_{+}(x_{\varphi},t)
p_{-}(x_{\varphi},t)\cos[\Omega(x_{\varphi},t)] \langle \alpha_{0} |
- \alpha_{0}  \rangle^{\eta(t)} \},
\label{Pgatto}
\end{equation}
where the first two terms,
\begin{equation}
 p^{2}_{\pm}(x_{\varphi},t)=\frac{1}{\sqrt{\pi \sigma^{(2)}(t)}} \, 
\exp\left\{-\frac{\big(x_{\varphi} \mp 
\mbox{Re}\{\alpha_{0}e^{-i\varphi}\} \chi(t)\big)^{2}} {\sigma^{(2)}(t)} \right\}
\label{gaussiane}
\end{equation}
pertain to the two initial coherent state, while the third, containing the functions
\begin{equation}
\Omega(x_{\varphi},t)=\frac{2 x_{\varphi} \mbox{ Im}\{\alpha_{0}e^{-i 
\varphi}\}\; \chi(t) }{\sigma^{(2)}(t)}
\label{omega}
\end{equation}
and 
\begin{equation}
\eta(t)=1- \frac{[\chi(t)]^{2}}{2 \sigma^{(2)}(t)},
\label{eta}
\end{equation}
describes the time evolution of the quantum interference between them. 
In Ref. \cite{GTV} it is shown that, in the case of zero feedback 
delay time ($\tau=0$) and perfect homodyne detection $\eta=1$, 
this interference term decays
with a decoherence time
\begin{equation}
t_{dec}(g) \simeq \frac{1}{2 \gamma |\alpha_{0}|^{2} (1- g 
\sin(\theta-\varphi))^{2}},
\label{tdec}
\end{equation}
which for $0\leq g \sin(\theta-\varphi) \leq 2$ implies that 
quantum coherence survives for a longer time with respect to the no-feedback
case ($g=0$).
If we consider the presence of a non-zero delay time, the correction 
to this feedback-induced decoherence slowing down can be evaluated 
from the behavior of the fringe visibility function $\eta(t)$ of 
Eq.~(\ref{eta}). It is however more instructive to see the effects of 
the feedback delay on 
the plots of the probability distribution. In Fig.~\ref{Pdix} and
Fig.~\ref{Pdixtfisso} we show the plots of $P(x_{\varphi},t)$ 
($x=x_{\varphi =0}$) for the 
case $\alpha_{0} = i5$. What is relevant in Fig.~2 is that the 
probability distribution in the presence of a non-zero delay 
$\gamma \tau =0.01$ and
efficiency $\eta=1$ (Fig.~2(c))
looses its interference fringes only 
slightly faster than the ideal case of zero delay and $\eta =1$ (Fig.~2(b)) 
and that the decoherence is still much slower than the no-feedback 
case (Fig.~2(a)). Moreover in Fig.~2(c) it is once again very evident the 
``step effect'', 
producing a rapid initial ``flattening'' of the probability 
distribution, which is quantitatively the main effect of the feedback delay.
Fig.~2(d) shows the effect of a non-unit detection efficiency (
$\eta 0.9$) which, as it can be easily expected, degrades the
performance of the homodyne feedback scheme in an appreciable way.

Fig.~2 and Fig.~3 show that an appreciable
decoherence retardation 
is obtained when the condition $\gamma\tau\leq 0.01$
is satisfied and this means a feedback delay time equal to
one half of the decoherence time in absence of feedback, 
$t_{dec}(0)=0.02 \gamma^{-1}$. 
Therefore, the feedback--induced decoherence retardation 
takes place even in the presence of a non-zero delay and Figs.~2 and 3 
show that one can even
tolerate delay times of the order of the decoherence time itself.

\subsection{Complete solution of the dynamics}

In this section we exactly solve the time evolution of the cavity mode 
in terms of the symmmetrically ordered
characteristic function, which is nothing but the expectation value of the 
cavity mode dispacement operator on the initial state of the whole system 
\begin{equation}
\chi_{W}(\lambda,t)={\rm Tr}\left(\rho_{T}(0) D(\lambda,t)\right)
\;\;\;\;\;\;\;\;\;
D(\lambda,t) = \exp\big( \,\lambda a^{\dagger}(t) - \lambda^{*} a(t) \, \big).
\label{d}
\end{equation}
Using the commutation rules of this operator with
$a(t)$, Eq.~(\ref{l1}) becomes in this case:
\begin{equation}
 \begin{array}{c}
   d D(\lambda,t)  =  \frac{ \gamma}{2} \Big\{ \lambda^{*} D(\lambda,t) a(t) - \lambda a^{\dagger}(t) D(\lambda,t) \Big\}
           dt + \: \sqrt{ \gamma} \, \lambda^{*}  D(\lambda,t) \, dB(t)\\  \\
   - \: \sqrt{ \gamma} \, \,\lambda  D(\lambda,t)
          dB^{\dagger}(t)   - \frac{\gamma}{8\eta} \, g^{2} ( \lambda e^{-i \theta} +  \lambda^{*} e^{i 
  \theta})^{2} D(\lambda,t) \Theta(t -
        \tau) dt\\ \\
  + i \, \frac{\sqrt{ \gamma}}{2} g ( \lambda e^{-i \theta} +  \lambda^{*} e^{i 
  \theta}) \, \Big\{ \: \sqrt{ \gamma} a^{\dagger}(t - \tau) \,e^{i \varphi}
      \,dt + \frac{dB_{f}^{\dagger}(t -\tau)}{\sqrt{\eta}} e^{i \varphi} 
      \: \Big\} \;D(\lambda,t)\, \Theta(t - \tau) \\  \\
  + i \, \frac{\sqrt{ \gamma}}{2} g ( \lambda e^{-i \theta} +  \lambda^{*} e^{i 
  \theta}) \,D(\lambda,t) \; \Big\{ \: \sqrt{ \gamma} a(t - \tau) \, e^{ - i
        \varphi} \,dt + \frac{dB_{f}(t - \tau)}{\sqrt{\eta}}  
        e^{ - i \varphi} \: \Big\} \,\Theta(t - \tau). 
 \end{array}
\label{equad}
\end{equation}
When we consider the initial condition of Eq.~(\ref{condin}),
we can simply focus on the matrix element
\begin{equation}
\langle D(\lambda,t)\rangle_{\mbox{{\tiny  $\beta \alpha$}}} 
= \Big\{ \langle 0 | \otimes \langle \beta | \Big\} \, D(\lambda,t) \, 
\Big\{ | \alpha \rangle \otimes | 0 \rangle \Big\},
\label{matrix}
\end{equation}
which obeys the following evolution equation 
\begin{equation}
   \frac{d}{d t} \langle D(\lambda,t)\rangle_{\mbox{{\tiny  $\beta \alpha$}}} \hspace{.2in}=
  \begin{array}[t]{c} 
   \frac{ \gamma}{2} \Big\{ \lambda^{*} \langle D(\lambda,t) a(t) \rangle_{\mbox{{\tiny  $\beta \alpha$}}}
   - \lambda \langle a^{\dagger}(t) D(\lambda,t) \rangle_{\mbox{{\tiny  $\beta \alpha$}}}\Big\} \\  \\
   - \frac{\gamma}{8\eta} \, g^{2} ( \lambda e^{-i \theta} +  \lambda^{*} e^{i 
  \theta})^{2}
   \langle D(\lambda,t)\rangle_{\mbox{{\tiny  $\beta \alpha$}}} \Theta(t - \tau) \\ \\
   + i \frac{\gamma}{2} g ( \lambda e^{-i \theta} +  \lambda^{*} e^{i 
  \theta})\, \Big\{ \: e^{i \varphi}
   \langle a^{\dagger}(t -\tau) \,D(\lambda,t) \rangle_{\mbox{{\tiny  $\beta \alpha$}}} \\ \\
   +  e^{ - i \varphi}  \langle D(\lambda,t) \, a(t - \tau) \rangle_{\mbox{{\tiny  $\beta \alpha$}}}
   \: \Big\} \,\Theta(t- \tau).
 \end{array}
\label{Dmedia}
\end{equation}
To solve Eq.~(\ref{Dmedia}), we have to deal with terms of the form
$\langle D(\lambda,t) \, a(t') 
\rangle_{\mbox{{\tiny  $\beta \alpha$}}}$. We first focus on its
$t'$ dependence, which can be determined in the same way as we have 
done in the previous Section for 
$\langle X_{\varphi}(t)  X_{\varphi}(t')\rangle_{\mbox{{\tiny $\beta 
\alpha$}}}$, that is, by
differentiating with respect to $t'$ $\langle D(\lambda,t) \, a(t') 
\rangle_{\mbox{{\tiny  $\beta \alpha$}}}$, by keeping $t$ constant. 
Using the commutation rules of Eq.~(\ref{neq5A}), 
it is possible to derive the following relations, 
valid for all positive value of $t$ and $t'$:
\begin{equation}
\begin{array}{lll}
	\big[ D(\lambda,t), a(0) \big] =D(\lambda,t) {\cal V}_{\lambda}(t) & & 
	\big[ D(\lambda,t), a^{\dagger}(0) \big] =D(\lambda,t) {\cal 
	V}_{\lambda}^{*}(t) \\ \\
	\big[ D(\lambda,t), dB(t') \big] = D(\lambda,t) {\cal R}_{\lambda}(t-t') dt' & &
\big[ D(\lambda,t), dB^{\dagger}(t') \big] = D(\lambda,t) {\cal 
R}_{\lambda}^{*}(t-t') dt', \\ \\
\big[ D(\lambda,t), dB_{f}(t') \big] = D(\lambda,t) {\cal 
R}_{\lambda}^{f}(t-t') dt' & &
\big[ D(\lambda,t), dB_{f}^{\dagger}(t') \big] = D(\lambda,t) {\cal 
R}_{\lambda}^{f\,*}(t-t') dt',
	\end{array}
\label{neqa30}
\end{equation}
where we have defined 
\begin{equation}
	{\cal V}_{\lambda}(t)=-\lambda e^{-\frac{\gamma}{2} t} - 
	\frac{i}{2}e^{i \varphi} \big(\,\lambda e^{-i \theta}+\lambda^{*} e^{i 
	\theta}\,\big) \Theta(t-\tau) \frac{\chi(t)-e^{-\frac{\gamma}{2} 
	t}}{\sin(\theta-\varphi)}
	\label{neq31}
\end{equation}
\begin{equation}
	{\cal R}_{\lambda}(t)=\lambda {\cal R}_{1}(t) - \lambda^{*}{\cal R}(t),
	\label{neq32}
\end{equation}
\begin{equation}
	{\cal R}_{\lambda}^{f}(t)=\lambda {\cal R}_{1}^{f}(t) - \lambda^{*}{\cal 
	R}^{f}(t),
	\label{neq32bis}
\end{equation}
(${\cal R}(t)$, ${\cal R}_{1}(t)$, ${\cal R}^{f}(t)$, and
${\cal R}_{1}^{f}(t)$ are given by Eqs.~(\ref{neq7B}), (\ref{neq7A}),
(\ref{neq7Bbis}) and (\ref{neq7Abis})).
Using Eqs.~(\ref{neqa30}), we then obtain 
\begin{eqnarray}
\langle D(\lambda,t) \, a(t') \rangle_{\mbox{{\tiny  $\beta 
\alpha$}}}&=&
\langle D(\lambda,t)  
\rangle_{\mbox{{\tiny  $\beta \alpha$}}} \Big\{ \alpha 
e^{-\frac{\gamma}{2} t'} -\frac{i}{2} e^{i \theta} \big( \beta^{*} 
e^{i \varphi} + \alpha e^{-i \varphi} \big) \Theta(t'-\tau) \frac{\chi(t')-e^{-\frac{\gamma}{2} 
	t'}}{\sin(\theta-\varphi)} \nonumber \\ \label{neq33} \\
	& & + e^{-\frac{\gamma}{2}( t'-\tau)}\Theta(t'-\tau) {\cal 
	W}_{\lambda}(t'-\tau,t) \Big\},\nonumber
	\end{eqnarray}
where 
\begin{equation}
{\cal W}_{\lambda}(t'-\tau,t)= -i\frac{\sqrt{\gamma}}{2} g e^{i \theta} 
	\int_{0}^{t'-\tau} dt^{\prime \prime} e^{\frac{\gamma}{2}t^{\prime 
	\prime}} \Big\{ 2 \sqrt{\gamma} \Lambda_{\lambda}(t^{\prime \prime},t) + 
	e^{i \varphi} \frac{{\cal R}_{\lambda}^{f\,*}(t-t^{\prime 
	\prime})}{\sqrt{\eta}} \Big\}
	\label{neq34}
\end{equation}
\begin{eqnarray}
\Lambda_{\lambda}(t,t')&=& -\frac{\sqrt{\gamma}}{2} e^{i\varphi} 
\int_{0}^{t} dt^{\prime \prime}\, \chi(t-t^{\prime \prime}) \Big[ \,
\Theta(t'-t^{\prime \prime}) {\cal R}_{\lambda}^{*}(t'-t^{\prime 
\prime}) - g \sin(\theta -\varphi)\Theta(t^{\prime \prime}-\tau) 
\frac{{\cal R}_{\lambda}^{f\,*}(t'-t^{\prime 
\prime}+\tau)}{\sqrt{\eta}} \, \Big]
\nonumber \\	\label{neq35}  \\
 & & + \frac{1}{2} e^{i \varphi} \chi(t) {\cal V}_{\lambda}^{*}(t'). 
 \nonumber
\end{eqnarray}
Using the fact that
$	\langle a^{\dagger}(t')\,D(\lambda,t) \rangle_{\mbox{{\tiny  $\beta 
\alpha$}}}= \langle D(-\lambda,t) \,a(t') \rangle_{\mbox{{\tiny  
$\alpha\beta$}}}^{*}$ and Eq.~(\ref{neq33}),
Eq.~(\ref{Dmedia}) 
becomes the simple homogeneous differential equation 
\begin{equation}
	   \frac{d}{d t} \langle D(\lambda,t)\rangle_{\mbox{{\tiny  $\beta 
	   \alpha$}}} = {\cal H}(t) \, \langle D(\lambda,t)\rangle_{\mbox{{\tiny  $\beta 
	   \alpha$}}}
	\label{neq37}
\end{equation}
whose solution is
\begin{equation}
	\langle D(\lambda,t)\rangle_{\mbox{{\tiny  $\beta 
	   \alpha$}}} = \langle D(\lambda,0)\rangle_{\mbox{{\tiny  $\beta 
	   \alpha$}}} \; \exp \Big[ \int_{0}^{t} dt^{\prime \prime} {\cal H}(t^{\prime 
	   \prime})\Big].
	\label{neq37A}
\end{equation}
This result is only apparently simple, since the explicit
time dependence of ${\cal H}(t)$ is given by
\begin{eqnarray}
	{\cal H}(t)&=& \frac{\gamma}{2} \big( \lambda^{*} \alpha - \lambda 
	\beta^{*} \big) e^{-\frac{\gamma}{2} t} + \frac{i}{2}\big( \lambda 
	e^{-i \theta} + \lambda^{*} e^{i \theta} \big) \big( \beta^{*} e^{i 
	\varphi} + \alpha e^{-i \varphi} \big) \Theta(t-\tau) \frac{ 
	\dot{\chi}(t) + \frac{\gamma}{2} e^{-\frac{\gamma}{2} t} 
	}{\sin(\theta -\varphi)}\nonumber \\ \nonumber \\
	& & - \frac{\gamma}{8\eta} g^{2} \big( \lambda 
	e^{-i \theta} + \lambda^{*} e^{i \theta} \big)^{2} \Theta(t-\tau) +
	 \frac{\gamma}{2} \Theta(t-\tau) e^{-\frac{\gamma}{2}(t-\tau)} 
	\Big[ \lambda^{*} {\cal W}_{\lambda}(t-\tau,t) + \lambda {\cal 
	W}_{\lambda}^{*}(t-\tau,t) \Big]\nonumber \\ 
	\label{neq38} \\
	& & +i \frac{\gamma}{2} g  \big( \lambda 
	e^{-i \theta} + \lambda^{*} e^{i \theta} \big) \Theta(t-2\tau) e^{-\frac{\gamma}{2}(t-2\tau)}
	\Big[ e^{-i\varphi} {\cal W}_{\lambda}(t-2\tau,t) -e^{i \varphi} {\cal 
	W}_{\lambda}^{*}(t-2\tau,t) \Big]\nonumber.
\end{eqnarray}
Eqs.~(\ref{neq37A}) and (\ref{neq38}) describe the time evolution of the 
cavity mode starting from the initial condition (\ref{condin}), in the 
case of a non-zero feedback delay time. It is however interesting to 
consider the approximated expression of this result at first order in 
$\gamma \tau$ since this condition can be easily realized 
experimentally with usual electro-optical feedback loops and good 
cavities.

\subsection{Approximated expression for $\langle D(\lambda,t)\rangle_{\mbox{{\tiny  $\beta 
\alpha$}}}$ in the $\gamma \tau \ll 1$ limit}

We have two possible equivalent ways to deal with the $\gamma \tau \ll 
1$ limit. The most straightforward one is simply to consider this limit
in the exact solution Eq.~(\ref{neq37A}). However this procedure is 
not very trasparent from the physical point of view
because of the complicated form of the 
function ${\cal H}(t)$. It is instead more instructive to perform the same limit 
from the beginning on 
the evolution equation for $\langle D(\lambda,t)\rangle_{\mbox{{\tiny  $\beta 
\alpha$}}}$, Eq.~(\ref{Dmedia}), and then integrate it. We 
shall consider this second approach also because it can be adopted 
not only in the problem considered in this paper (linear 
choice (\ref{qfb}) for the feedback operator $F(t)$)
but also for more general forms of the operator $F(t)$. 
Let us go back therefore to Eq.~(\ref{Dmedia}), where the difficult 
terms to handle are those containing $\langle D(\lambda,t) \, a(t - \tau) \rangle_{\mbox{{\tiny 
$\beta \alpha$}}}$ and its complex conjugate, which can be rewritten as
\begin{equation}
\langle D(\lambda,t) \, a(t - \tau) \rangle_{\mbox{{\tiny  $\beta \alpha$}}} = 
\langle D(\lambda,t) \, a(t)
\rangle_{\mbox{{\tiny  $\beta \alpha$}}} - \langle D(\lambda,t) \; \Delta a(t - \tau) \rangle_{\mbox{{\tiny 
$\beta \alpha$}}},
\label{approx1}
\end{equation}
where 
\begin{equation}
 \Delta a(t - \tau)\equiv a(t) - a(t - \tau).
\label{approx2}
\end{equation}
In the limit $\gamma \tau \ll 1$, we can use Eq.~(\ref{l1}) to approximate 
Eq.~(\ref{approx2}) at the first order in 
$\tau$ so that we can write (we also consider $t\geq 2 \tau$) 
\begin{equation}
\begin{array}{c}
\langle D(\lambda,t) \, a(t - \tau) \rangle_{\mbox{{\tiny  $\beta \alpha$}}} \simeq  
\langle D(\lambda,t) \, \Big\{ a(t)  + \frac{ \gamma}{2} \, a(t - \tau) \, \tau +  i g \gamma e^{i \theta} 
  X_{\varphi}(t- 2 \tau)\, \tau \Big\} \rangle_{\mbox{{\tiny  $\beta \alpha$}}}\\  \\  
  + \: i \, g\,\sqrt{\frac{\gamma}{4\eta}}\, e^{i(\theta + \varphi)}\:\langle D(\lambda,t) \;
  \Delta B_{f}^{\dagger}(t- 2 \tau) \rangle_{\mbox{{\tiny  $\beta \alpha$}}},
\end{array}
\label{approx3} 
\end{equation}
where $\Delta B_{f}(t- 2 \tau)$ is the following Ito increment
\begin{equation}
\Delta B_{f}(t- 2 \tau) = B_{f}(t-\tau)-B_{f}(t-2\tau). 
\label{deltaa1}
\end{equation}
The last term in the right side of 
Eq.~(\ref{approx3}) can be simplified using the identity
\begin{equation}
D(\lambda,t) = D(\lambda,t-2\tau) + \Big\{ \; D(\lambda,t) - 
D(\lambda,t-2\tau) \; \Big\}
\label{approx4}
\end{equation}
and approximating the term in the curly brackets again at first order 
in $\tau$. 
Finally one gets
\begin{eqnarray}
\lefteqn{\langle D(\lambda,t) \, a(t - \tau) \rangle_{\mbox{{\tiny  $\beta \alpha$}}} \simeq
 \Big( 1 + \frac{ \gamma \tau}{2} (1 + i g  e^{i (\theta - \varphi)}) \, \Big) \langle D(\lambda,t) \,a(t) 
\rangle_{\mbox{{\tiny  $\beta \alpha$}}} } \nonumber \\ 
\label{approx6} \\
& &+ \, i \frac{ \gamma \tau}{2}  e^{i(\theta + \varphi)}\langle a^{\dagger}(t) \, D(\lambda,t) \rangle_{\mbox{{\tiny  $\beta
\alpha$}}}  - \frac{ \gamma \tau}{4 \eta}  g^{2} (\lambda e^{-i \theta} + 
\lambda^{*} e^{i \theta}) e^{i \theta}  \langle D(\lambda,t) \rangle_{\mbox{{\tiny 
$\beta \alpha$}}}.  \nonumber 
\end{eqnarray}
Replacing this approximated expression in Eq.~(\ref{Dmedia}) 
together with the corresponding one for
 $\langle a^{\dagger}(t -\tau) \,D(\lambda,t) \rangle_{\mbox{{\tiny  $\beta
\alpha$}}}$, one obtains a simple integrable equation (valid for
$t\geq2\tau$) having the following solution:
\begin{equation}
\langle D(\lambda,t)\rangle_{\mbox{{\tiny  $\beta \alpha$}}}= 
\langle \beta | \alpha \rangle \, \exp\left\{ - {\cal A}(t)|\lambda|^{2} + {\cal B}_{1}(t) \lambda^{2}
+{\cal B}_{2}(t) \big(\lambda^{*}\big)^{2} + {\cal C}(t) \lambda + {\cal D}(t) \lambda^{*} \right\}
\label{Dapprox}
\end{equation}
where 
\begin{eqnarray}
&&{\cal A}(t)=\frac{1}{2} + \frac{g^{2}}{4\eta} \Big\{ {\cal F}_{0}(t) - \gamma \tau \,{\cal F}_{1}(t) \Big\}\\
\nonumber\\ &&{\cal B}_{1}(t)=\Big({\cal B}_{2}(t)\Big)^{*}= 
-\frac{g^{2}}{8\eta} \Big\{ {\cal F}_{0}(t) - \gamma \tau \,{\cal
F}_{1}(t) \Big\} \, e^{-2 i \theta},
\end{eqnarray}
in which
\begin{equation}
{\cal F}_{0}(t) = \frac{1- e^{-(1-2 g \sin\!\theta)\gamma t}}{1-2 g \sin\!\theta} 
\end{equation}
\begin{equation}
{\cal F}_{1}(t) = (1+ \gamma t\,g \sin\!\theta) e^{-(1-2 g \sin\!\theta)\gamma t} + (g\, \sin\!\theta) {\cal F}_{0}(t), 
\end{equation}
and 
\begin{eqnarray}
{\cal C}(t)&=&-\frac{i}{2 \sin\!\theta}( \alpha e^{-i \theta} + \beta^{*} e^{i \theta}) e^{-\frac{\gamma}{2}t} +
\frac{ie^{-i\theta}}{2 \sin\!\theta}( \alpha  + \beta^{*}) e^{-(1-2 g \sin\!\theta)\frac{\gamma}{2}t} \nonumber \\ 
\nonumber \\
& &+\left\{\frac{i}{2}ge^{-i \theta} ( \alpha  + \beta^{*}) ((1- 2 g \sin\!\theta)\frac{\gamma}{2} t-1)e^{-(1-2 g
\sin\!\theta)\frac{\gamma}{2}t} \right\}\, \gamma \tau, \\ \nonumber \\
{\cal D}(t)&=&-\frac{i}{2 \sin\!\theta}( \alpha e^{-i \theta} + \beta^{*} e^{i \theta}) e^{-\frac{\gamma}{2}t} +
\frac{ie^{i\theta}}{2 \sin\!\theta}( \alpha  + \beta^{*}) e^{-(1-2 g \sin\!\theta)\frac{\gamma}{2}t} \nonumber \\ 
\nonumber \\
& &+\left\{\frac{i}{2}ge^{i \theta} ( \alpha  + \beta^{*}) ((1- 2 g \sin\!\theta)\frac{\gamma}{2} t-1)e^{-(1-2 g
\sin\!\theta)\frac{\gamma}{2}t} \right\}\, \gamma \tau.
\end{eqnarray}
(We have chosen the phases so that 
$\varphi=0$). As expected, setting $\tau=0$ one obtains the same results 
of Ref. \cite{GTV}. 
It is important to note that Eq.~(\ref{Dapprox}) has been obtained 
integrating from $t=2\tau$, i.e., using $\langle 
D(\lambda,2 \tau)\rangle_{\mbox{{\tiny  $\beta \alpha$}}}$ as initial 
condition. This initial condition could be derived from the exact 
solution Eq.~(\ref{neq37A}), but it could also be obtained
by noting that Eq.~(\ref{Dmedia}) takes a very simple form
for $0 \leq t \leq 2\tau$. In fact the terms 
$\langle D(\lambda,t) \, a(t - \tau) \rangle_{\mbox{{\tiny  $\beta \alpha$}}}$
and $\langle a^{\dagger}(t -\tau) \,D(\lambda,t) \rangle_{\mbox{{\tiny  
$\beta \alpha$}}}$, for $\tau \leq t \leq 2 \tau$, can be written as
\begin{eqnarray}
	\langle D(\lambda,t) \, a(t - \tau) \rangle_{\mbox{{\tiny  $\beta 
	\alpha$}}}&=& \alpha \langle D(\lambda,t) \rangle_{\mbox{{\tiny  $\beta 
	\alpha$}}} e^{-\frac{\gamma}{2}(t-\tau)} \nonumber \\ 
	\label{neq40}  \\ 
	\langle a^{\dagger}(t -\tau) \, D(\lambda,t) \rangle_{\mbox{{\tiny  $\beta 
	\alpha$}}}&=& \beta^{*} \langle D(\lambda,t) \rangle_{\mbox{{\tiny  $\beta 
	\alpha$}}} e^{-\frac{\gamma}{2}(t-\tau)}, \nonumber 
\end{eqnarray}
depending on the fact that for $\tau \leq t \leq 2 \tau$ the 
operators $a(t-\tau)$ and $a^{\dagger}(t-\tau)$ are not affected by the 
feedback loop. Moreover, for $0 \leq t \leq \tau$ these terms do not contribute 
to the evolution because of the presence of the step function in the equation. 
In this way Eq. (\ref{Dmedia}) can be simply integrated for $0 \leq t \leq 
2\tau$ too. 

\begin{figure}
\centerline{\epsfig{figure=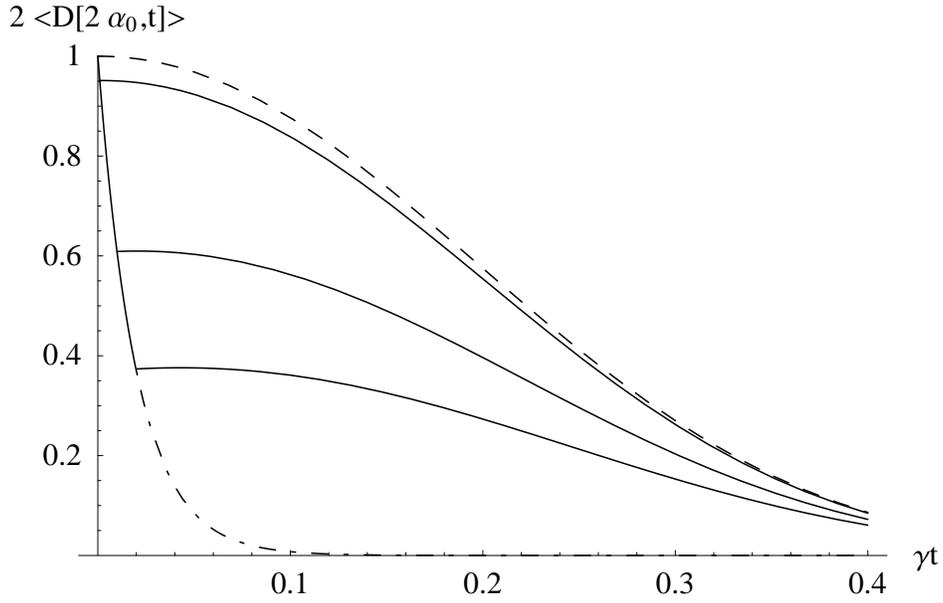,width=13cm}}
\vspace{0.2cm}
\caption{Time evolution of $2 \langle D(2 \alpha_{0},t) \rangle$ 
for $\alpha_{0}=i5$ and $\varphi=0$
for $g\sin\!\theta=0$ (dot-dashed
line), $g\sin\!\theta=1$ with $\gamma \tau=0$ (dashed line) 
and $g\sin\!\theta=1$ with, from bottom to top,
 $\gamma \tau=0.02$, $\gamma \tau=0.01$, $\gamma \tau=0.001$ (full lines).}
\label{coerenza}
\end{figure}

In Ref.~\cite{GTV}, the decoherence inhibition capabilities of the 
homodyne-mediated feedback scheme have been described by looking at 
the so-called coherence function $C(t)=\langle -\alpha_{0} |\rho(t)|
\alpha_{0} \rangle$ . However, an equivalent description of the 
decoherence of the
 Schr\"{o}dinger-cat state (\ref{gatto1}) is provided by 
the time evolution of the expectation value of the operator $D(2 
\alpha_{0},t)$ on the state (\ref{gatto1}). In fact, for 
$|\alpha_{0}|^{2} \gg 1$ and
$\gamma t \ll 1$, this quantity has the same 
time behavior of the coherent function $C(t)$ of Ref.~\cite{GTV}, 
because its off-diagonal contributions are 
greater then the diagonal ones by the exponential factor
$e^{2|\alpha_{0}|^{2}}$. In the small delay time limit $\gamma \tau \ll 1$,  
this equivalent coherence function takes a simple form, given by
\begin{equation}
\begin{array}{c}
\langle D(2 \alpha_{0},t) \rangle \simeq \frac{1}{2} \exp\!\Big\{ - 2|\alpha_{0}|^{2} \Big[
2 + \frac{(g\sin\!\theta)^{2}}{\eta}({\cal F}_{0}(t) - \gamma \tau \,{\cal F}_{1}(t) ) \\ \\
 -2 e^{-\frac{\gamma}{2}(1- 2 g \sin\!\theta) t} 
 + (g \sin\!\theta) \big[ 2 - \gamma t (1 - 2 g \sin\!\theta)\big] 
 \gamma \tau e^{-\frac{\gamma}{2}(1- 2 g \sin\!\theta)
t}\Big] \Big\},
\end{array} 
\label{coerenza2}
\end{equation}
where we have defined phases so that $\varphi = 0$
and considered ${\rm Re}\{\alpha_0\}=0$.
In the zero delay time case ($\tau =0$), when $\gamma t\ll 1$, this 
function is an
exponential with characteristic time equal to (\ref{tdec}) (see 
Ref.~\cite{GTV}). 
We have plotted Eq.~(\ref{coerenza2}) in Fig.~\ref{coerenza}
again for $\alpha_{0}=i5$ in the ideal case $\eta=1$
for different values of the feedback delay
and we have compared it with the no-feedback case (dot-dashed line) 
and with the ideal case $\tau=0$ and $\eta=1$ (dashed line). 
As we have seen above, the decoherence slowing down is significant up 
to $\gamma \tau =0.01$, i.e. $\tau =0.5 t_{dec}$
and of course increases as long as the feedback delay decreases.
In Fig.~5 the effect of an imperfect homodyne detection is considered 
and it is shown how, as expected, the decay of the coherence function $D(2 
\alpha_{0},t)$ becomes faster and faster as long as the efficiency
$\eta$ decreases.

\section{Conclusions}

In this paper the dynamics of a cavity mode subject to a feedback loop 
using a fraction of the output homodyne photocurrent to control 
the transmittivity of a mirror has been completely solved in the
general non-Markovian case in which the feedback delay time is not
negligible. To solve this problem we have generalized the derivation 
of the feedback quantum Langevin equation based on the input-output
theory of Ref.~\cite{wis3} to the non-unit detection efficiency case.
We have seen that when $\eta \leq 1$, the fed-back photocurrent
involves a new Ito noise term, coinciding with the usual input noise
only in the $\eta =1$ limit.

We have seen that the main effect of the delay is the 
``step effect'', that is, the fact that the feedback loop starts acting 
only after the time $\tau$ has elapsed. Apart from this, the dynamical behavior 
in the presence of a feedback delay does not differ very much from the 
predictions of the Markovian treatment based on the zero-delay limit, 
as long as $\gamma \tau \ll 1$.
As a consequence, the significant decoherence slowing down demonstrated
in Ref.~\cite{GTV} for the zero delay limit holds even in the presence
of a non-zero delay $\tau$, and one gets a good decoherence inhibition for
values of $\tau$ up to one half of the decoherence time.

\acknowledgments

This work has been partially supported by INFM (through the 1997 
Advanced Research Project ``Cat''), by the European Union
in the framework of the TMR Network ``Microlasers and Cavity QED'',
and by MURST through ``Cofinanziamento 1997''.

\appendix
\section{}

Let us consider the evolution equation for the cavity mode
state vector $| \psi_{w}(t) \rangle$ for the case of feedback with zero
delay time, given in
\cite{GTV}:
\begin{equation}
d |\psi_{w}(t) \rangle =\{ A dt + B dw(t) \}| \psi_{w}(t) \rangle,
\label{j1}
\end{equation}
in which  $dw(t)$ is a real-valued Wiener increment deriving from the input noise, and 
\begin{eqnarray}
A&=&-\frac{\gamma}{2} a^{\dagger}a - \frac{\gamma}{2} F^{2} -i\gamma F a e^{-i\varphi}, \label{j2} \\ \nonumber \\
B&=&\sqrt{\gamma} ( a e^{-i\varphi} -i F), \label{j3} 
\end{eqnarray}
are cavity mode operators in the Schr\"{o}dinger picture, with \mbox{$F=g (a e^{-i \theta} + a^{\dagger} e^{i \theta})/2$} as
in Eq. (\ref{qfb}): this equation can be derived from Eq. (\ref{GTV1}) 
by means of the method developed in \cite{gg1}.  
Now we adopt the stochastic integration method derived in \cite{jac} to explicitely 
solve Eq. (\ref{j1}). The explicit analytical solution is given by
\begin{equation}
| \psi_{w}(t) \rangle=E_{w}(t) | \psi_{w}(0) \rangle,
\label{j4}
\end{equation}
where the evolution operator is
\begin{equation}
E_{w}(t)= \exp\big[i \Lambda(t)\big]\;\exp\big[(A-\frac{B^{2}}{2}) \,t\big]\;\exp\big[\chi_{-}(t) a +
\chi_{+}(t) a^{\dagger}\big],
\label{j5}
\end{equation}
with $\Lambda(t)$ and $\chi_{\pm}(t)$ complex functionals of the Wiener 
process $w(t)$; defining in particular the
functions 
\begin{eqnarray}
f_{1}(t)\!= \!\frac{\sqrt{\gamma}}{2^{3/2}}\Big\{\!\!-ig(1\!+e^{-i2 \varphi})e^{i\theta} e^{\frac{\gamma}{2}t}
 + \big[\!-ige^{-i\theta} + 2e^{-i\varphi} + i g e^{i (\theta - 2\varphi)}\big] e^{-\frac{\gamma}{2}t} \Big\} \label{f1} \\
\nonumber \\
f_{2}(t)\!=\! \frac{\sqrt{\gamma}}{2^{3/2}}\Big\{\!\!-g(1\!-e^{-i2 \varphi})e^{i\theta} e^{\frac{\gamma}{2}t}
 + i\big[\!-ige^{-i\theta} + 2e^{-i\varphi} + i g e^{i (\theta - 2\varphi)}\big] e^{-\frac{\gamma}{2}t} \Big\}
\label{f2}
\end{eqnarray}
we have that 
\begin{eqnarray}
\chi_{j}(t) = \int_{0}^{t} f_{j}(t') dw(t')    \hspace{0.5 in} \mbox{for   } j = 1,2 \label{j6} \\ \nonumber \\
\Lambda(t)= \int_{0}^{t} f_{1}(t')\chi_{2}(t') dw(t') -\int_{0}^{t} f_{2}(t') \chi_{1}(t')dw(t') \label{j7} \\ \nonumber \\
\chi_{\pm}(t)= \frac{1}{\sqrt{2}} \big( \chi_{1}(t) \pm i \chi_{2}(t) \big) \label{j8}.
\end{eqnarray}
If we consider the case of an initial coherent state $|\alpha_{0} 
\rangle$, Eq.~(\ref{j5}) gives
\begin{equation}
| \psi_{w}(t) \rangle = {\cal E}_{w}(t) | [ \chi_{+}(t) + \alpha_{0}] 
e^{-\frac{\gamma}{2}t} \rangle
\label{j9}
\end{equation}
that is, the state remains coherent with amplitude 
$[ \chi_{+}(t) + \alpha_{0}] e^{-\frac{\gamma}{2}t}$, 
and ${\cal E}_{w}(t)$ is a complex
weight given by
\begin{equation}
{\cal E}_{w}(t)\;=
\begin{array}[t]{l}
 \exp\!\Big\{i \Lambda(t) + \frac{1}{2} (\chi_{+}(t)^{*} + \chi_{-}(t)) (\chi_{+}(t) + 2 \alpha_{0}) + i \mbox{
Im}[\chi_{+}(t)\alpha_{0}^{*}] \\ \\
+\, i\frac{g}{4} \gamma t e^{i(\theta-\varphi)} -\frac{1}{2} \big[|\chi_{+}(t) + \alpha_{0}|^{2}+ (\chi_{+}(t) +
\alpha_{0})^{2} e^{-2i\varphi}\big] (1-e^{-\gamma t}) \Big\}.
\end{array}
\label{inutile}
\end{equation}
First of all we note that replacing $g=0$ in Eq. (\ref{j9}) we correctly obtain 
the same results of Ref. \cite{gg2} for
the case of no feedback loop. The state remains coherent in agreement 
with the "no-go" theorem of Ref. \cite{wis2}, according to which
homodyne-mediated feedback is not able to increase the nonclassicality of the output light. 
Using linearity, from Eq.(\ref{j9}), one easily gets the evolution of the 
Schr\"{o}dinger-cat state
\begin{equation}
| \psi_{Cat} \rangle =  N\Big\{| \alpha_{0} \rangle \: + \: |- \alpha_{0}  \rangle  \Big\}
\label{gattoj}
\end{equation}
and it is also possible to evaluate the coherent function defined in \cite{GTV},
\begin{equation}
C_{w}(t)= \frac{\langle -\alpha_{0} | \psi_{w}(t) \rangle \langle  \psi_{w}(t) | \alpha_{0} \rangle}{\langle  \psi_{w}(t) |
\psi_{w}(t) \rangle }.
\label{j11}
\end{equation}
By considering the limits $|\alpha_{0}| \gg 1$ and $\gamma t \ll 1$ 
and setting set $\varphi=0$ and $\mbox{Re}\{\alpha_{0}\}=0$ as 
in \cite{GTV}, one gets the following expression
\begin{equation}
C_{w}(t)=\frac{1}{2} \exp\!\Big\{ -\frac{1}{4} \gamma g^{2} w^{2}(t) - 2 i \sqrt{\gamma} |\alpha_{0}| (1-g \sin\!\theta)
w(t) \Big\},
\label{j12}
\end{equation}
reproducing very well the numerically obtained stochastic 
trajectories of Ref. \cite{GTV}. 
Eq. (\ref{j12}) clearly shows the results of Ref.~\cite{GTV}, i.e. 
that the modulus of
$C_{w}(t)$ is not $\theta$-dependent and that the large fluctuations 
in the absence of feedback and those in the presence of feedback but
with phase $\theta \neq  \pi/2$ are essentially phase fluctuations.

\section{}

Let us consider the equation (\ref{qfi}) for the function 
$\langle X_{\varphi}(t) \rangle_{\mbox{{\tiny  AB}}}$,
\begin{equation}
  \frac{d}{d \xi} Z(\xi) = -  Z(\xi) + 2 k \, Z(\xi - y)\, \Theta(\xi - y),
\label{zeta}
\end{equation}
where for sake of simplicity we have introduced the following notation:
\begin{equation}
 \begin{array}{c}
\xi = \frac{\gamma t}{2}, \hspace{.4 in} y = \frac{\gamma \tau}{2} \\ \\
 Z(\xi) = \langle X_{\varphi}(\frac{2 \xi}{\gamma}) \rangle_{\mbox{{\tiny  AB}}}\\ \\
 k = g \sin( \theta - \varphi). 
 \end{array}
\label{costantiadim}
\end{equation}

We are interested to study Eq. (\ref{zeta}) for $\xi \geq 0$, with initial condition
\mbox{$ Z(0) = Z_{0}$}.
First of all we observe that $Z(\xi)$ is a continuous function of class $C^{\infty}(I)$ 
on every interval 
$I = \big] \, n \, y, (n+1) \, y \, \big[ $ \/, with \mbox{$n \in N$}. 
Moreover it is possible to verify 
that its $p^{th}$ derivative is not continuous at $ \xi = py $.
Apart from this irregular behaviour 
due to the presence of $ \Theta(\xi - y)$, 
it is easy to see from Eq. (\ref{zeta}) that 
the generic solution $Z(\xi)$ is bounded by a locally integrable function 
and therefore it can be Laplace-transformed.
Denoting with $\tilde{Z}(s)$ the Laplace transform of
$Z(\xi)$, from Eq. (\ref{zeta}) we have
\begin{equation}
\tilde{Z}(s) = \frac{Z_{0}}{(s+1)\big( 1 - \frac{2 k \, e^{- s y}}{s+1}\big)}.
\label{laplace1}
\end{equation}
It is always possible to choose the integration path for the antitransformation so
that 
\begin{equation}
\left| \frac{2 k \, e^{- sy}}{s+1} \right| < 1,
\label{maggio}
\end{equation}
and using the geometrical series, we can write (\ref{laplace1}) as
\begin{equation}
\tilde{Z}(s) = Z_{0} \sum_{n=0}^{+ \infty}\frac{(2 k )^{n}\, e^{-s n y}}{(s+1)^{n+1}}
\label{llaplace}
\end{equation}
which can be easily antitransformed.
Using relations Eq. (\ref{costantiadim}) we 
then obtain the solution (\ref{soluzione}) of Section 2.2.

\begin{figure}
\centerline{\epsfig{figure=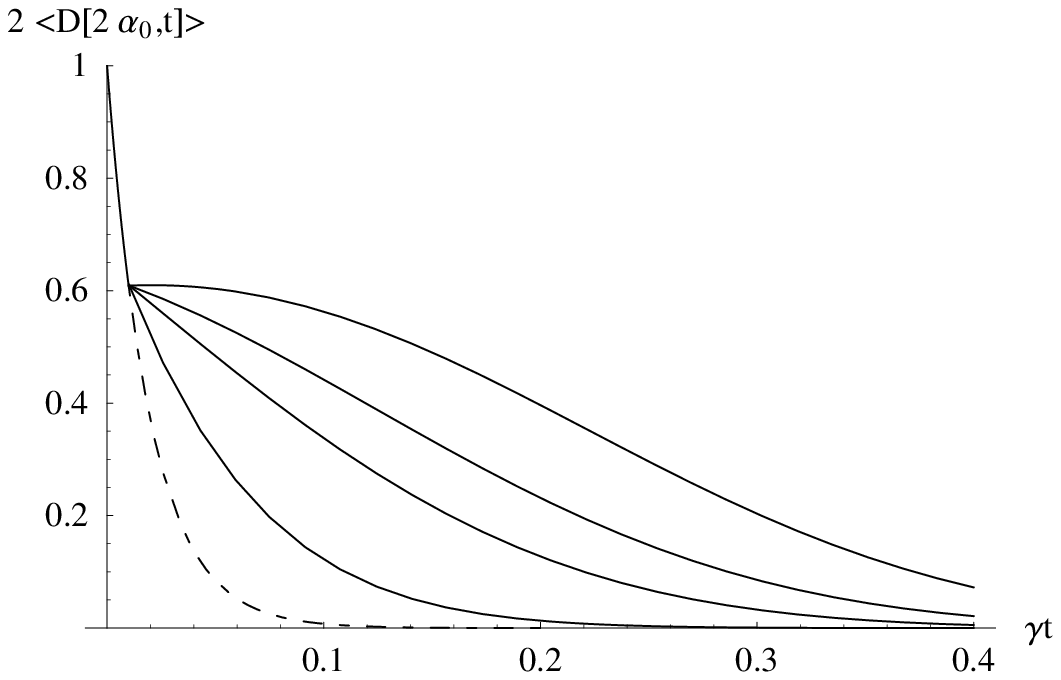,width=13cm}}
\vspace{0.2cm}
\caption{Time evolution of $2 \langle D(2 \alpha_{0},t) \rangle$ 
for $\alpha_{0}=i5$ and $\varphi=0$ in the presence of feedback
with non-zero delay time ($\gamma \tau =0.01$ and $g\sin\!\theta=1$) 
and for different values of the homodyne detection efficiency
$\eta$; from bottom to top,
$\eta=0.75$, $\eta=0.9$, $\eta=0.95$, $\eta=1$ (full lines).
The dashed
line refers to the no-feedback case.}
\label{coerenzaeta}
\end{figure}


\begin{thebibliography}{99}


\bibitem{wis1}H.M. Wiseman and G.J. Milburn, Phys. Rev. Lett. {\bf 70}, 548 (1993).
\bibitem{wis2}H.M. Wiseman and G.J. Milburn, Phys. Rev. A {\bf 49}, 1350 (1994).
\bibitem{wis3}H.M. Wiseman, Phys. Rev. A {\bf 49}, 2133 (1994).
\bibitem{squee}P. Tombesi and D. Vitali, Phys. Rev. A {\bf 50}, 4253 (1994).
\bibitem{qnd}P. Tombesi and D. Vitali, Appl. Phys. B {\bf 60}, S69 (1995);
Phys. Rev. A {\bf 51}, 4913 (1995).
\bibitem{GTV}P. Goetsch, P. Tombesi and D. Vitali, Phys.
Rev. A {\bf 54}, 4519 (1996).
\bibitem{prl}D. Vitali, P. Tombesi, G.J. Milburn, Phys. Rev. Lett. 
{\bf 79}, 2442 (1997).
\bibitem{opto}S. Mancini, D. Vitali, and P. Tombesi, Phys. Rev. Lett.
{\bf 80}, 688 (1998).
\bibitem{pran}D. Vitali, P. Tombesi, G.J. Milburn, Phys. Rev. A 
{\bf 57}, 4930 (1998).
\bibitem{poy}J.F. Poyatos, J.I. Cirac and P. Zoller, Phys. Rev. Lett. 
{\bf 77}, 4728 (1996).
\bibitem{eke}A. Ekert and R. Josza, Rev. Mod. Phys {\bf 68}, 733 (1996).
\bibitem{molm}B.M. Bay, P. Lambropoulos, and K. Molmer, Phys. Rev. Lett.
{\bf 79}, 2654 (1997).
\bibitem{diosi}L. Di\'osi, N. Gisin, and W.T. Strunz, Phys. Rev. A 
{\bf 58}, 1699 (1998); T. Yu, L. Di\'osi, N. Gisin, and W.T. Strunz, 
LANL e-print archive quant-ph/9902043.
\bibitem{jack}M.W. Jack, M.J. Collett, and D.F. Walls, 
LANL e-print archive quant-ph/9807028.
\bibitem{carm}H.J. Carmichael, {\it An Open System Approach to Quantum Optics},
Lecture Notes in Physics m18, (Springer, Berlin, 1993).
\bibitem{gar0}M.J. Collett and C.W. Gardiner, Phys. Rev. A {\bf 30}, 1386 (1984); 
C.W. Gardiner and M.J. Collett, Phys. Rev. A
{\bf 31}, 3761 (1985).
\bibitem{gar1}C.W. Gardiner {\em Quantum Noise} (Springer, Berlin, 1991).
\bibitem{gar2}C.W. Gardiner, A.S. Parkins and P. Zoller, Phys. Rev. A {\bf 46}, 4363 (1992).
\bibitem{gg1}P. Goetsch and R. Graham, Phys. Rev. A {\bf 50}, 5242 (1994).
\bibitem{homo}H.M. Wiseman and G.J. Milburn, Phys. Rev. A {\bf 47}, 
642 (1993).
\bibitem{jac}K. Jacobs and P.L. Knight, Phys. Rev. A {\bf 57}, 2301 (1998).
\bibitem{gg2}P. Goetsch, R. Graham and F. Haake, Phys. Rev. A {\bf 51}, 136 (1995).

\end{thebibliography}
\end{document}